\documentclass[12pt]{article}
\usepackage{verbatim,color,setspace,graphicx,epsfig,rotating, epstopdf,lscape}
\usepackage{amsmath,amsfonts, lineno, hyperref, geometry, bbm, environ,censor,calc} 
\usepackage[utf8]{inputenc}
\usepackage[english]{babel}
\usepackage{comment}
\usepackage{pdfcolfoot}
\usepackage{natbib} %standard package for references
\usepackage[title]{appendix}
\usepackage[normalem]{ulem} % used for \sout{}
\definecolor{orange}{rgb}{0.8, 0.33, 0.0}
\definecolor{green}{rgb}{0.0, 0.42, 0.24}

\newcommand{\blind}{0}

\NewEnviron{mysout}
 {\soutrefunexp{\BODY}}
\newlength\nextcharwidth
\makeatletter
\renewcommand\@cenword[1]{%
 \setlength{\nextcharwidth}{\widthof{#1}}%
 \censorrule{\nextcharwidth}%
 \kern -\nextcharwidth%
 #1}
\makeatother
\newcommand\soutref[1]{\censorruledepth=.55ex\xblackout{#1}}
\newcommand\soutrefunexp[1]{\expandafter\soutref\expandafter{#1}}

\begin{document}

\def\spacingset#1{\renewcommand{\baselinestretch}%
{#1}\small\normalsize} \spacingset{1}

\if0\blind
{
\title{Forecasting with Economic News}
\author{Luca Barbaglia$^{a}$, Sergio Consoli$^{a}$, Sebastiano Manzan$^{b}$\\
\mbox{}\\
{\small $^{a}$ Joint Research Centre, European Commission}\\
{\small $^{b}$ Zicklin School of Business, Baruch College}\\
\\ \mbox{} \\
}}\fi
\if1\blind
{
 \title{Forecasting with Economic News}
 \author{
 \mbox{}\\
 \mbox{}\\
 \mbox{}\\ 
 \mbox{} \\
 }
} \fi

%\date{\today}
\date{}
\maketitle
\thispagestyle{empty}
\setcounter{page}{1}

% \vspace{-0.5cm}
% \begin{center}
% \Large{\textsc{Preliminary and incomplete: \\ please do not circulate}}
% \end{center}

\vspace{0.2cm}
\begin{abstract}
\begin{center}
\mbox{}\\
\begin{minipage}{15cm}
\noindent
{\small 
The goal of this paper is to evaluate the informational content of sentiment extracted from news articles about the state of the economy. We propose a fine-grained aspect-based sentiment analysis that has two main characteristics: 1) we consider only the text in the article that is semantically dependent on a term of interest (aspect-based) and, 2) assign a sentiment score to each word based on a dictionary that we develop for applications in economics and finance (fine-grained). 
Our data set includes six large US newspapers, for a total of over 6.6 million articles and 4.2 billion words. 
Our findings suggest that several measures of economic sentiment track closely business cycle fluctuations and that they are relevant predictors for four major macroeconomic variables. 
We find that there are significant improvements in forecasting when sentiment is considered along with macroeconomic factors. 
In addition, we also find that sentiment matters to explains the tails of the probability distribution across several macroeconomic variables. 
 } 
\mbox{}\\
\mbox{}
\mbox{}
\end{minipage}
\end{center}
\end{abstract}

\if0\blind
{
\vspace{1cm}
\begin{center}
\begin{minipage}{15cm}
\singlespacing
{\footnotesize The views expressed are purely those of the authors and do not, in any circumstance, be regarded as stating an official position of the European Commission. We are grateful to participants of the ``Alternative data sets for Macro Analysis and Monetary Policy'' conference held at Bocconi University, the 
``Nontraditional Data \& Statistical Learning with Applications to Macroeconomics” Banca d’Italia and Federal Reserve Board joint conference, and seminar participants at the Bank of Spain, University of Amsterdam, Universidad Autonoma de Madrid and Maastricht University for numerous comments that significantly improved the paper. We are also greatly indebted to the Associate Editor and Referees for insightful comments and to the Centre for Advanced Studies at the Joint Research Centre for the support, encouragement, and stimulating environment while working on the {\it bigNOMICS} project. {\tt E-mail}: luca.barbaglia@ec.europa.eu, sergio.consoli@ec.europa.eu, and sebastiano.manzan@baruch.cuny.edu.
% \\ 
% $^*$ Corresponding author: Luca Barbaglia.
}
\end{minipage}
\end{center}
}\fi
\if1\blind
{ }\fi

\newpage

\spacingset{1.84} 

\section{Introduction}
\label{sec:introduction}

Economic forecasts are an essential input to design fiscal and monetary policies, and improving their accuracy is a continued challenge for economists. A reason for the often disappointing performance of economic forecasts is that any individual indicator provides a very noisy measure of the state of the economy. In the last 20 years, several methods have been proposed to extract a robust signal from large panels of macroeconomic variables \citep[see][for a comprehensive review]{stock2016dynamic}. In addition, many economic variables are available only monthly or quarterly and are released with considerable delay, which complicates the monitoring and forecasting of the economy in real-time. This has led to the development of nowcasting methods that capitalize on the continuous flow of economic information released by statistical agencies \citep[see][]{banbura2013now}. 
\cite{bok2018macroeconomic} provide a recent and updated discussion of the state of economic forecasting.

%\bigskip\noindent
A promising path to increase both the accuracy and the timeliness of economic forecasts is to use alternative data sets to complement the information provided by the statistical agencies. 
By ``alternative" we refer to data sets collected as the outcome of a business transaction (e.g., credit card or online purchases), tracking the media (e.g., news or twitter), or internet searches (e.g., google trends) among others. 
What these data sets have in common is that, in most cases, they are collected in real-time by companies rather than being produced by government agencies through surveys. 
The researcher is then able to aggregate the granular information in these data sets to build indicators that can be used in forecasting models. 
Realistically, these alternative indicators should be expected to complement, rather than replace, the macroeconomic variables provided by the government agencies. 

%\bigskip \noindent
On the one hand, the official statistics are very accurate measures of an economic concept of interest (e.g., industrial production or inflation) and the outcome of a well-designed and time-tested sampling strategy. 
Nevertheless, the fact that these variables are obtained from business and consumer surveys entails an infrequent sampling period (typically, monthly or quarterly), and publication delays deriving from the collection and processing of the answers. 
On the other hand, indicators based on alternative data sets, in most cases, provide a biased sample of the population and might not measure accurately a specific economic variable. 
However, they might be available in real-time which makes them very appealing to monitor the state of the economy. 
Hence, the trade-off between accuracy of the predictors and their real-time availability summarises the potential added-value of alternative indicators relative to the official statistics provided by government agencies. 
A recent example is \cite{lewis2020us}, who construct a weekly indicator of economic activity pooling information from raw steel production, fuel sales, road traffic, and electricity output among others. Other alternative data sets used for nowcasting and forecasting macroeconomic variables are credit card transactions \citep[see][]{galbraith2018nowcasting, aprigliano2019using}, google trends \citep[][]{choi2012predicting}, road tolls \citep{askitas2013nowcasting} or firm-level data \citep{fornaro2016predicting}.

%\bigskip\noindent
In this paper we use news from six major US outlets as our alternative data set and construct predictors that measure sentiment about different aspects of economic activity. 
More specifically, we identify the text in the articles published in a certain day that contains a token (i.e., term) of interest (e.g., {\it economy} or {\it inflation}). We then consider the words that are semantically dependent on the token of interest and use only those words to calculate the sentiment measure. 
Our sentiment analysis is thus local in nature, in the sense that it considers only the text related to a term of interest, as opposed to a global approach that evaluates the whole article. The benefit of our methodology is that it provides a more accurate measure of the sentiment associated to an economic concept, instead of measuring sentiment on a large body of text that might involve and mix different economic concepts. 
In addition, even when topic analysis is used to cluster articles, sentiment is typically calculated by pooling the text of all the articles in the cluster, which might involve the discussion of different economic variables. 
Another contribution of this paper is that we develop a dictionary that we specifically construct having in mind applications in the economic and financial domains. 
Similarly to \cite{loughran2011liability}, the dictionary contains words that are frequently used in economics and finance
and we assign a sentiment value between $\pm$ 1 rather than categorizing the terms in positive and negative. 
Then, we use this dictionary to assign a sentiment value to the words related to the term of interest. 
We refer to our approach as Fine-Grained Aspect-based Sentiment (FiGAS) which highlights the two main ingredients of the approach: a value between $\pm$ 1 for the sentiment carried by a word based on our dictionary (fine-grained) and the sentiment calculated in a neighborhood of a token of interest and only on semantically related words (aspect-based). 

%% EXAMPLE
%\bigskip\noindent
As an illustration, consider the situation in which we are interested to measure the sentiment in the news about the overall state of the economy. 
In this case we could specify {\it economy} as our token of interest and identify the following sentence\footnote{The sentence was published in the Wall Street Journal on 11/02/2016.}: ``{\it Her comments played into the concern that after years of uneven growth, the U.S. economy is becoming more vulnerable to the global slowdown}''. The term of interest {\it economy} is related to the verb {\it become} whose meaning is modified by the adjective {\it vulnerable} which is further modified by the adverb {\it more}\footnote{Notice that \textit{become}, \textit{more} and \textit{vulnerable} are directly associated to the term of interest, while \textit{global slowdown} is indirectly related through the dependence with {\it vulnerable}.}.
In terms of sentiment, our dictionary assigns a negative value to {\it vulnerable} and a positive one to {\it more}. Hence, the sentiment associated with the term {\it economy} in this sentence will be negative and larger (in absolute value) relative to the sentiment of {\it vulnerable} due to the contribution of the adverb {\it more}. 
This example also shows that the approach is interpretable in the sense that it provides the researcher with the information determining the sentiment value relative to alternative approaches, for instance based on machine learning models, that do not provide a narrative \citep[][]{thorsrud2020words}. 

%% LITERATURE
%\bigskip\noindent
The interest of economists in using text, such as news, is growing. 
An early application is \citet{baker2016measuring}, who construct a measure of economic and political uncertainty based on counting the number of articles that contain tokens of interest. 
Other papers aim at measuring economic sentiment conveyed by news to explain output and inflation measures for the US \citep[][]{shapiro2020measuring,kelly2018text,bybee2019structure}, the UK \citep[][]{kalamaramaking}, and Norway \citep[][]{thorsrud2016nowcasting, thorsrud2020words}. 
More generally, economists are interested in analyzing text as an alternative source of data to answer long standing questions \citep[][]{gentzkow2017text}, such as the role of central bank communication \citep[][]{hansen2016shocking,hansen2017transparency}, asset pricing \citep[][]{calomiris2019news}, economic expectations \citep[][]{lamla2012role,sharpe2017s,ke2019predicting}, stock market volatility \citep[][]{baker2019policy}, and media bias \citep[][]{gentzkow2010drives} among others. 
Overall, these papers show that news provide information relevant to a wide range of applications, 
although a relevant question is why that is the case. \cite{blinder2004does} argue that the media is an important source of information for consumers about the state of the economy. 
News reported by television and newspapers are an essential input in the formation of expectations and in consumption and investment decisions. 
In addition, news provide information about a dispersed set of events, such as recently released data, monetary and fiscal decisions, domestic and international economic events, but also provide views about the past, current, and future economic conditions formulated by economists, policy-makers, and financial analysts.

%\bigskip\noindent
%% OUR APPLICATION
In our empirical application we construct sentiment measures based on over 6.6 million news articles that we use as predictors to forecast the quarterly real GDP growth rate and three monthly variables, namely the growth rate of the Industrial Production Index and the Consumer Price Index, and the change in non-farm payroll employment. 
All variables enter the model in real-time, that is, we use all available information up to the date at which we produce the pseudo forecasts.
The indicators are designed to capture the sentiment related to a single token of interest ({\it economy}, {\it unemployment}, {\it inflation}) or to a group of tokens of interest, such as {\it monetary policy} (e.g., central bank, federal funds, etc), {\it financial sector} (e.g., banks, lending, etc), and {\it manufacturing} (e.g., manufacturing, industrial production, etc). 
In addition, we refine the FiGAS approach discussed earlier along two dimensions. 
First, we identify the prevalent geographic location, if any, of the article. Given our goal of forecasting US macroeconomic variables we exclude from the calculation of sentiment the articles that refer to the rest of the world. Second, we detect the verbal tense of the text related to the token of interest. 

%% RESULTS
Our results are as follows. First, we find that our economic sentiment measures have a strong correlation with the business cycle, in particular for the {\it economy}, {\it unemployment} and {\it \textcolor{black}{manufacturing}} measures. 
In addition, the sentiment about the {\it financial sector} and {\it monetary policy} shows pessimism during recessions and optimism during expansions, although they tend to fluctuate also in response to other events. 
In particular, we find that the tendency of the {\it financial sector} indicator to be negative during recessions reached dramatic lows during the Great Recession and was followed by a very slow recovery. So, it is encouraging to see that, qualitatively, the proposed measures seem to capture the different phases of the cycle and provide an explanation for the relevance of each variable to the business fluctuations.
Second, augmenting a factor-based model with economic sentiment measures delivers higher forecast accuracy.
The predictive advantage of sentiment is present also at the longer horizons considered when the macro-factors are less relevant. Instead, at the nowcasting and backcasting horizons the flow of macroeconomic information carries most of the predictive power for nearly all variables. 
An encouraging result of the analysis is that some of the sentiment measures provide systematically higher accuracy, even in the case of a difficult variable to forecast, such as CPI inflation. 
In particular, we find that the sentiment about the {\it economy} is very ery often the most relevant predictor for GDP and non-farm payroll, the indicators about \textit{unemployment} and \textit{monetary policy} helps predicting industrial production, while the {\it financial sector} sentiment is important in forecasting the growth rate of CPI.

%% QUANTILES
We also consider the effect of economic sentiment at the tails of the distribution by extending the regression models to the quantile settings. Our results show that sentiment indicators become more powerful predictors when considering the tails of the distribution. In particular, for the monthly variables, we find that indicators related to {\it economy}, {\it financial sector} and {\it monetary policy} contribute significantly to increase forecast accuracy relative to the available macroeconomic information. This is a favorable result that could provide more accurate measures of Growth-at-Risk (see \citealp{adrian2018term} and \citealp{brownlees2021backtesting} for further discussion about Growth-at-Risk applications).

%\bigskip\noindent
Concluding, there is encouraging evidence to suggest that economic sentiment obtained from news using the FiGAS approach captures useful information in forecasting macroeconomic variables such as quarterly GDP and monthly indicators. 
These results show that analyzing text from news can be a successful strategy to complement the macroeconomic information deriving from official releases. 
In addition, the availability of news in real-time allows the high-frequency monitoring of the macroeconomic variables of interest. The paper is organized as follows: in Section \ref{sec:sec1} we introduce the FiGAS approach, the measures of economic sentiment, and the forecasting models.
Section \ref{Sec_data} describes the data set.
We then continue in Sections \ref{sec:in_sample} and \ref{sec:out_of_sample} with the discussion of the in-sample and out-of-sample results, respectively. Finally, Section \ref{sec:sec3} concludes.

\section{Methodology}
\label{sec:sec1}

The application of text analysis in economics and finance has, in most cases, the goal of creating an indicator that measures the sentiment or the intensity of a certain topic in the news. 
Despite this common goal, the literature presents a wide and diverse range of techniques used in the analysis. 
%% NEWS SELECTION
The first difference among the approaches relates to the way news are selected to compute the indicator. 
An early approach followed by \cite{tetlock2007} is to use a column of the Wall Street Journal as the relevant text source, while discarding the rest of the news. 
An alternative approach is followed by \cite{baker2016measuring}, who select a subset of articles that contain tokens of interest related to economic and political uncertainty. 
Several recent papers \citep[see][among others]{kelly2018text, thorsrud2016nowcasting, thorsrud2020words} rely on topic analysis, such as the Latent Dirichlet Allocation (LDA), which represents an unsupervised modelling approach to clustering articles based on their linguistic similarity. The analysis of the most frequent words within each cluster provides then insights on the topic. 

%% SENTIMENT COMPUTATION
The second difference regards the measure used to compute sentiment or intensity in the selected text. 
One simple measure that has been used is to count the number of articles that contain the tokens of interest \citep[see][]{baker2016measuring} or the number of articles belonging to a certain topic \citep[see][]{kelly2018text}. 
The goal of such measure is to capture the time-varying attention in the media for a certain topic, although it does not take into account the tone with which the news talk about the topic. 
An alternative approach is to compute a measure of sentiment which accounts for the positive/negative tone of a word. 
This is typically done by counting the words that are positive and negative in the text based on a dictionary \citep[see][for an early application in finance]{tetlock2007}. 
A sentiment measure is then given by the difference between the number of positive and negative words, standardized by the total number of words in the article. The advantage of this approach is that it measures the strength and the direction of the message reported by the news on a certain topic which is likely to have an impact on the economic decisions of agents. \cite{algaba2020} is an extensive survey that reviews the recent developments in the application of sentiment analysis to economics and finance. 
In the following section we describe in general terms our approach to sentiment analysis and how we use these measures to forecast four main macroeconomic indicators. More details on the methodology are provided in Appendix \ref{Appendix_FiGAS}. 

\subsection{Fine-grained aspect-based sentiment analysis}

%There is a vast literature on Sentiment Analysis (SA) with applications ranging from psychology, linguistics to economics and finance.\cite{algaba2020} provide an updated overview of the recent developments in this field, with particular focus on econometric applications. They propose the following generic definition of sentiment as \textit{``the disposition of an entity towards an entity, expressed via a certain medium"}. We identify three key elements in this definition: (i) the entity expressing the disposition, (ii) the entity about which the disposition is referring to, (iii) the disposition expressed via the medium. In this paper, we use sentiment analysis to quantify how (i) the major US outlets presented in Section \ref{Sec_data} talk about (ii) a specific aspect of the economy, using a (iii) continuous score in the [-1,+1] interval as a measure of the expressed sentiment polarity, with +1 and -1 being, respectively, the higher and lower bounds for the sentiment scores. 

%The FiGAS approach to sentiment analysis proposed by \cite{recupero2015sentilo} is based on two key elements. 
Our FiGAS approach leverages on recent advances in sentiment analysis \citep{sentic2015cambria,Cambria2020105,Xing201849}. It is based on two key elements. The first is that it is {\it fine-grained} in the sense that words are assigned a polarity score that ranges between $\pm$1 based on a dictionary. 
The second feature is that it is {\it aspect-based}, that is, it identifies chunks of text in the articles that relate to a certain concept and calculates sentiment only on that text, rather than the full article.
To relate this approach to the existing literature, similarly to \cite{baker2016measuring} we identify a subset of the text that relates to the aspect or token of interest. 
In addition, similarly to \cite{tetlock2007} the sentiment is based on assigning a value or a category (positive/negative) based on a dictionary instead of simply counting the number of articles. A novel aspect of our analysis is that we use a fine-grained dictionary that has not been used previously for applications in economics and finance. 

%\bigskip\noindent
The first step of the analysis consists of creating a list of tokens of interest (ToI)\footnote{To find an exhaustive list of terms on a certain economic concept we rely on the World Bank Ontology available at: \url{http://vocabulary.worldbank.org/thesaurus.html}.} that collects the terms that express a certain economic concept and for which we want to measure sentiment. 
In our application, we construct six economic indicators based on the following ToI:
\begin{itemize}
 \item {\it economy}: economy;
 \item {\it financial sector}: bank, derivative, lending, borrowing, 
 real estate, equity market, stock market, bond, insurance rate, exchange rate and combinations of [banking, financial] with [sector, commercial, and investment];
 \item {\it inflation}: inflation; 
 \item {\it \textcolor{black}{manufacturing}}: manufacturing and combinations of [industrial, manufacturing, construction, factory, auto] with [sector, production, output, and activity]; 
 \item {\it monetary policy}: central bank, federal reserve, money supply, monetary policy, federal funds, base rate, and interest rate; 
 \item {\it unemployment}: unemployment.
\end{itemize}
The selection of topics and ToI is made by the researcher based on the application at hand. 
In making the choice of ToIs, we were driven by the goal of measuring news sentiment that would be predictive of economic activity. 
Hence, we designed the ToIs in order to capture text that discusses various aspects of the overall state of the {\it economy}, the {\it inflation} rate, the {\it unemployment} rate, the banking and {\it financial sector}, {\it \textcolor{black}{manufacturing}}, and {\it monetary policy}. Given that our application is focused on forecasting US macroeconomic variables, we only selects articles that do not explicitly mention a nation other than the US. 

%% FiGAS
%\bigskip\noindent
After selecting a chunk of text that contains one of our ToIs, we continue with the typical Natural Language Processing (NLP) workflow to process and analyze the text\footnote{For all our NLP tasks we use the \textit{spaCy} Python library and rely on the \textit{en\_core\_web\_lg} linguistic model. }:

\begin{itemize}
	\item \textit{Tokenization \& lemmatization}: the text is split into meaningful segments (\textit{tokens}) and words are transformed to their noninflected form (\textit{lemmas});
	%\item \textit{Named Entity Recognition}: Named-entity mentions in the news text are located and classified, including person names, locations, organizations, time expressions, quantities, monetary values, companies and product, percentages, etc. such as the person names, organizations, locations, medical codes, time expressions, quantities, monetary values, percentages, etc.
	\item \textit{Location detection}: we detect the most frequent location that has been identified in the text, if any; 
 \item \textit{Part-of-Speech (POS) tagging}: the text is parsed and tagged (see the Appendix for details on the POS);
 \item \textit{Dependency Parsing}: after a ToI is found in the text, we examine the syntactic dependence of the neighbouring tokens with a rule-based approach (see the Appendix for more details);
 \item \textit{Tense detection}: we detect the tense of the verb, if any, related to the ToI;
 \item \textit{Negation handling}: the sentiment score is multiplied by $-1$ when a negation term is detected in the text.
\end{itemize}

\noindent
The core of the FiGAS approach are the semantic rules that are used to parse the dependence between the ToI and the dependent terms in the neighboring text. Once these words are identified, we assign them a sentiment score which we then aggregate to obtain a sentiment value for the sentence. These sentiment scores are obtained from a fine-grained dictionary that we developed based on the dictionary of \cite{loughran2011liability} and additional details are provided in the Appendix. Each score is defined over $[-1,1]$ and sentiment is propagated to the other tokens such that the overall tone of the sentence remains defined on the same interval. We then obtain a daily sentiment indicator for the topic by summing the scores assigned to the selected sentences for that day. Hence, our sentiment indicator accounts both for the volume of articles in each day as well as for the sentiment. The resulting sentiment indicator is thus \textit{aspect-based}, since it refers to a certain topic of interest, and \textit{fine-grained}, since it is computed using sentiment scores defined on $[-1,1]$.

%% TENSE
%\bigskip\noindent

An additional feature of the proposed algorithm is the possibility of detecting the verbal tense of a sentence. To shorten the presentation we build our sentiment indicators aggregating over all tenses and report the results for the sentiment of the individual verbal tenses in Appendix \ref{sec_verbal}. 
We refer to \cite{consoli2021fine} for a detailed technical presentation of the proposed algorithm and for a comparison with other sentiment analysis methods that have been presented in the literature.
Finally, Python and R packages that perform the FiGAS analysis are available on the authors' personal websites.

\subsection{Forecasting models\label{sec_models}}
{\color{black}
The goal of our analysis is to produce forecasts for the value of a macroeconomic variable in period $t$ released in day $d$, which we denote by $Y_t^d$. 
We introduce the index $d$ to track the real-time information flow available to forecasters, and to separate it from the index $t$ that represents the reference period of the variable (e.g., at the monthly or quarterly frequency). 
As we discuss in Section \ref{Sec_data} below, in the empirical application we keep track of the release dates of all the macroeconomic variables included in the analysis. This allows us to mimic the information set that is available to forecasters on the day the forecast was produced. We indicate by $X_{d-h}$ the vector of predictors and lags of the dependent variable that are available to the forecaster on day $d-h$, that is, $h$ days before the official release day $d$ of the variable. 
%% UMIDAS
Since in our application the vector $X_{d-h}$ includes variables at different frequencies, we use the Unrestricted MIDAS (U-MIDAS) approach proposed by \cite{marcellino2010factor} and \cite{foroni2015unrestricted}, that consists of including the weekly and monthly predictors and their lags as regressors. Hence, $X_{d-h}$ includes an intercept, lagged values of the variable being forecast, and current and past values of weekly and monthly macroeconomic and financial indicators. The forecasting model that we consider is given by: 
\begin{equation}
Y_t^d = \eta_h S_{d-h} + X_{d-h}'\beta_{h} + \epsilon_{t,d-h},
\label{eq:double_lasso_equation1}
\end{equation}
where $X_{d-h}$ is defined above, $S_{d-h}$ is a scalar representing a %scalar 
sentiment measure available in day $d-h$, $\eta_h$ and $\beta_h$ are parameters to be estimated, and $\epsilon_{t,d-h}$ is an error term. 
The parameter $\eta_h$ represents the effect of a current change in the sentiment measure on the future realization of the macroeconomic variable, conditional on a set of control variables.
Notice that the model in Equation \eqref{eq:double_lasso_equation1} can also be interpreted in terms of the local projection method of \cite{jorda2005estimation}.

%% LASSO
In our application, we include a large number of predictors that might negatively affect the accuracy of the estimate due to the limited sample size available for quarterly and monthly variables. 
In order to reduce the number of regressors to a smaller set of relevant predictors, we use the \textit{lasso} penalization approach proposed by \cite{tibshirani1996regression}. 
The selected predictors are then used to estimate Equation~(\ref{eq:double_lasso_equation1}) 
in a post-lasso step that corrects the bias introduced by the shrinkage procedure and allows to assess the statistical significance of the sentiment measures. 
Unfortunately, this post-single lasso procedure produces inconsistent estimates due to the likelihood of omitting some relevant predictors from the final model. 
%% DOUBLE-LASSO
To account for this issue, we select the most important regressors relying on the data-driven \textit{double lasso} approach proposed in \cite{belloni2014inference}. This approach delivers consistent estimates of the parameters and is likely to have better finite sample properties which is especially important given the moderate sample size in the current application. 
The post double lasso procedure is implemented in the following two steps.
First, variables are selected using lasso selection in Equation \eqref{eq:double_lasso_equation1}, but also from a regression of the target variable, in our case the sentiment measure $S_{d-h}$ on $X_{d-h}$, that is:
\begin{equation}
S_{d-h} = X_{d-h}'\delta_{h} + \xi_{d-h}.
\label{eq:double_lasso_equation2}
\end{equation}
In the second step, the union of the variables selected in the two lasso first stage regressions are used to estimates Equation \eqref{eq:double_lasso_equation1} by OLS. \cite{belloni2014inference} shows the consistency and the asymptotic normality of the estimator of $\eta_h$. 
% In our application, we use the data-driven penalty proposed by \cite{belloni2014inference} to select the most important regressors.

%% QUANTILES
The focus on the conditional mean disregards the fact that predictability could be heterogeneous at different parts of the conditional distribution. This situation could arise when a sentiment indicator might be relevant to forecast, for instance, low quantiles rather than the central or high quantiles. 
There is evidence in the literature to support the asymmetric effects of macroeconomic and financial variables in forecasting economic activity, and in particular at low quantiles \citep[see][among others]{manzan2013macroeconomic,manzan2015forecasting,adrian2019vulnerable}. 
To investigate this issue we extend the model in Equation \eqref{eq:double_lasso_equation1} to a quantile framework: 
\begin{equation}
Q_{\tau,h}(Y_t^d) = \eta_{\tau,h} S_{d-h} + X_{d-h}'\beta_{\tau, h},
\label{eq:double_lasso_quantile1}
\end{equation}
where $Q_{\tau,h}(Y_t^d) $ is the $\tau$-level quantile forecast of $Y_t^d$ at horizon $h$.
The lasso penalized quantile regression proposed by \cite{koenker2011additive} can be used for variable selection in Equation \eqref{eq:double_lasso_quantile1}, followed by a post-lasso step to eliminate the shrinkage bias. 
Similarly to the conditional mean case, the post-lasso estimator of $\eta_{\tau,h}$ could be inconsistent due to the possible elimination of relevant variables during the selection step. 
Following \cite{belloni2019valid}, we implement the \textit{weighted double selection} procedure to perform robust inference in the quantile setting.
%\cite{belloni2019valid} propose two approaches to perform robust inference in the quantile setting that is characterized by the additional complication of the non-differentiability of the quantile loss function. 
%We implement their \textit{weighted double selection} approach proposed in \cite{belloni2019valid} which consists of the following steps. 
In the first step the method estimates Equation \eqref{eq:double_lasso_quantile1} using the lasso penalized quantile regression estimator. 
The fitted quantiles are then used to estimate the conditional density of the error at zero which provides the weights for a penalized lasso weighted mean regression of $S_{d-h}$ on $X_{d-h}$. 
In the second step, the union of the variables selected in the first stage regressions are used to estimate the model in Equation \eqref{eq:double_lasso_quantile1} weighted by the density mentioned earlier. 
The resulting $\eta_{\tau, h}$ represents the post double lasso estimate at quantile $\tau$ and it is asymptotically normally distributed. More details are available in \cite{belloni2019valid}.

%% ARX and ARXS
In the empirical application, the baseline specification includes lags of the dependent variable and three macroeconomic factors that summarize the variation across large panels of macroeconomic and financial variables. We denote this specification as $ARX$, and by $ARXS$ the ARX specification augmented by the sentiment measures. 
}

% \section{Application}
% \label{sec:sec2}

%\subsection{Data\label{Sec_data}}
\section{Data\label{Sec_data}}

Our news data set is extracted from the Dow Jones Data, News and Analytics (DNA) platform\footnote{Dow Jones DNA platform accessible at \url{https://professional.dowjones.com/developer-platform/}.}. We obtain articles for six newspapers\footnote{The New York Times, Wall Street Journal, Washington Post, Dallas Morning News, San Francisco Chronicle, and the Chicago Sun-Times. The selection of these newspapers should mitigate the effect of media coverage over time on our analysis since they represent the most popular outlets that cover the full sample period from 1980 to 2019.}
from the beginning of January 1980 until the end of December 2019 and select articles in all categories excluding sport news\footnote{More specifically we consider articles that are classified by Dow Jones DNA in at least one of the categories: economic news (ECAT), monetary/financial news (MCAT), corporate news (CCAT), and general news (GCAT).}. 
The data set thus includes 6.6 million articles and 4.2 billion words. The information provided for each article consists of the date of publication, the title, the body of the article, the author(s), and the category. 
%% MACRO
As concerns the macroeconomic data set, we obtain real-time data from the Philadelphia Fed and from the Saint Louis Fed ALFRED repository\footnote{Saint Louis Fed ALFRED data repository accessible at \url{https://alfred.stlouisfed.org/}.}, that provides the historical vintages of the variables, including the date of each macroeconomic release. 
The variables that we include in our analysis are: real GDP (GDPC1), Industrial Production Index (INDPRO), total Non-farm Payroll Employment (PAYEMS), Consumer Price Index (CPIAUCSL), the Chicago Fed National Activity Index (CFNAI), and the National Financial Conditions Index (NFCI). 
The first four variables represent the object of our forecasting exercise and are available at the monthly frequency, except for GDPC1 that is available quarterly. The CFNAI is available monthly and represents a diffusion index produced by the Chicago Fed that summarizes the information on 85 monthly variables \citep[see][]{brave2014nowcasting}. 
Instead, the NFCI is a weekly indicator of the state of financial markets in the US and is constructed using 105 financial variables with a methodology similar to CFNAI \citep[see][]{brave2009chicago}. 
In addition, we also include in the analysis the ADS Index proposed by \cite{aruoba2009real} that is maintained by the Philadelphia Fed. ADS is updated at the daily frequency and it is constructed based on a small set of macroeconomic variables available at the weekly, monthly and quarterly frequency. In the empirical application we aggregate to the weekly frequency by averaging the values within the week. 
We will use these three macroeconomic and financial indicators as our predictors of the four target variables. The ADS, CFNAI, and NFCI indicators provide a parsimonious way to include information about a wide array of economic and financial variables that proxy well for the state of the economy\footnote{An issue with using the three indexes is their real-time availability in the sample period considered in this study. In particular, the methodology developed by \cite{stock1999forecasting} was adopted by the Chicago Fed to construct the CFNAI index and, much later, the NFCI \citep{brave2011monitoring}. The ADS Index was proposed by \cite{aruoba2009real} and since then it is publicly available at the Philadephia Fed. Since we start our out-of-sample exercise in 2002, forecasters had only available the CFNAI in real-time while the ADS and NFCI indexes were available only later. 
Although this might be a concern for the real-time interpretability of our findings, we believe there are also advantages in using them in our analysis. The most important benefit is that the indexes provide a parsimonious way to control for a large set of macroeconomic and financial variables that forecasters observed when producing their forecast that are correlated with the state of the economy and our news sentiment.}. To induce stationarity in the target variables, we take the first difference of PAYEMS, and the percentage growth rate for INDPRO, GDPC1 and CPIAUCSL.

\begin{figure}[ht]
 \centering
 \includegraphics[scale = 0.5]{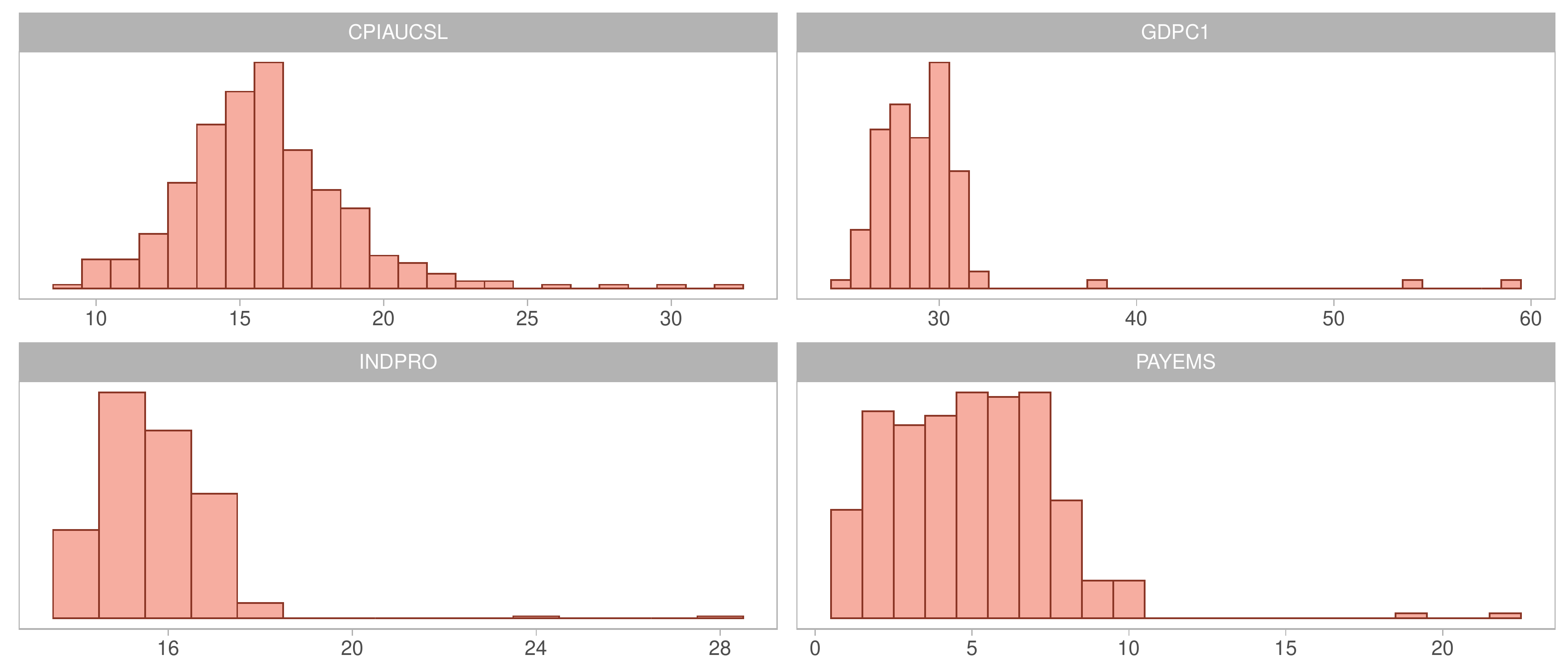}
 \caption{Histogram of the number of days between the end of the reference period and the date of the first release of the dependent variables, namely CPIAUCSL, GDPC1, INDPRO and PAYEMS.}
 \label{fig:pub_lag}
\end{figure}

%\bigskip\noindent
The release dates provided by ALFRED are available since the beginning of our sample in 1980 with two exceptions. 
For GDPC1, vintages are available since December 1991 and for CFNAI only starting with the May 2011 release. Figure \ref{fig:pub_lag} shows the publication lag for the first release of the four variables that we forecast. The monthly flow of macroeconomic information begins with the Bureau of Labor Statistics employment report that includes PAYEMS, typically in the first week of the month. During the second and third week of the month, CPIAUCSL and INDPRO are announced. Finally, the advance estimate of real GDP is released toward the end of the month following the reference quarter. 
We verified the dates of the outliers that appear in the figure and they correspond to governmental shutdowns that delayed the release beyond the typical period. Based on the vintages available since May 2011, the release of CFNAI typically happens between 19 and 28 days after the end of the reference period, with a median delay of 23 days. 
For the sample period before May 2011, we assign the 23rd day of the month as the release date of the CFNAI. This assumption is obviously inaccurate as it does not take into account that the release date could happen during the weekend and other possible patterns (e.g., the Friday of the third week of the month). In our empirical exercise the release date is an important element since it keeps track of the information flow and it is used to synchronize the macroeconomic releases and the news arrival. However, we do not expect that the assumption on the CFNAI release date could create significant problems in the empirical exercise. This is because the release of CFNAI is adjusted to the release calendar of the 85 monthly constituent variables so that no major releases in monthly macroeconomic variables should occur after the 23rd of the month.

\subsection{Economic sentiment measures}

\begin{figure}[ht]
 \centering
 \includegraphics[scale = 0.5]{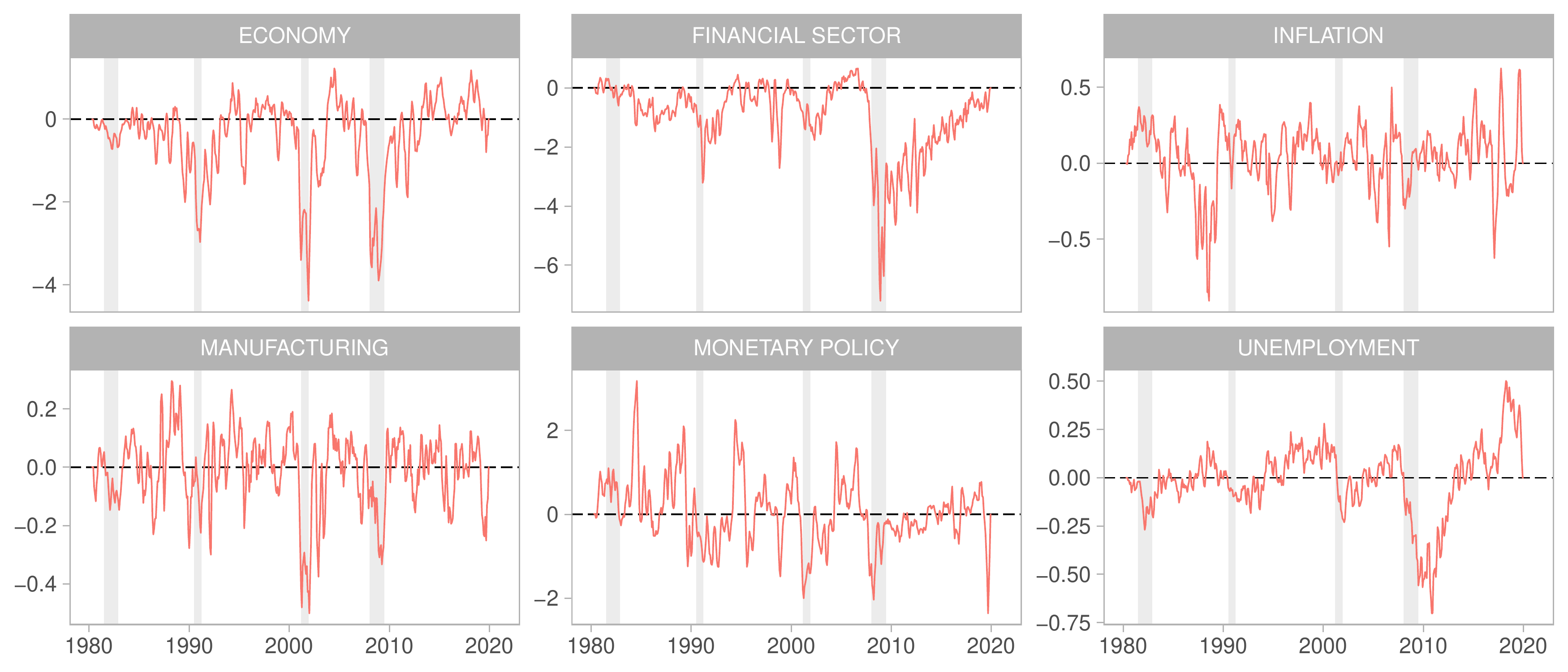}
 \caption{Time series of the sentiment measures for all verbal tenses. The daily series is smoothed by taking a moving-average of 30 days and then sub-sampled at monthly frequencies. The grey areas denote the NBER recessionary periods.}
 \label{fig:ts_sentiment_paprfunan}
\end{figure}

Figure \ref{fig:ts_sentiment_paprfunan} shows the six time series of the economic sentiment sampled at the monthly frequency when we consider all verbal tenses, together with the NBER recessions. 
The sentiment measures for {\it economy} and {\it manufacturing} seem to broadly vary with the business cycle fluctuations and becoming more pessimistic during recessions. 
The {\it unemployment} indicator follows a similar pattern, although the 2008-2009 recession represented a more extreme event with a slow recovery in sentiment. 
The measure for the {\it financial sector} indicates that during the Great Recession of 2008-2009 there was large and negative sentiment about the banking sector.
%% VOLUME SENTIMENT
As mentioned earlier, our sentiment measure depends both on the volume of news as well as on its tone. 
The financial crisis was characterized by a high volume of text related to the financial sector which was mostly negative in sentiment. The uniqueness of the Great Recession as a crisis arising from the financial sector is evident when comparing the level of sentiment relative to the values in earlier recessionary periods. The sentiment for {\it monetary policy} and {\it inflation} seems to co-vary to a lesser extent with the business cycle. In particular, negative values of the indicator of monetary policy seem to be associated with the decline in rates as it happens during recessionary periods, while positive values seems to be associated with expansionary phases. 

\begin{figure}[ht]
 \centering
 \includegraphics[scale = 0.5]{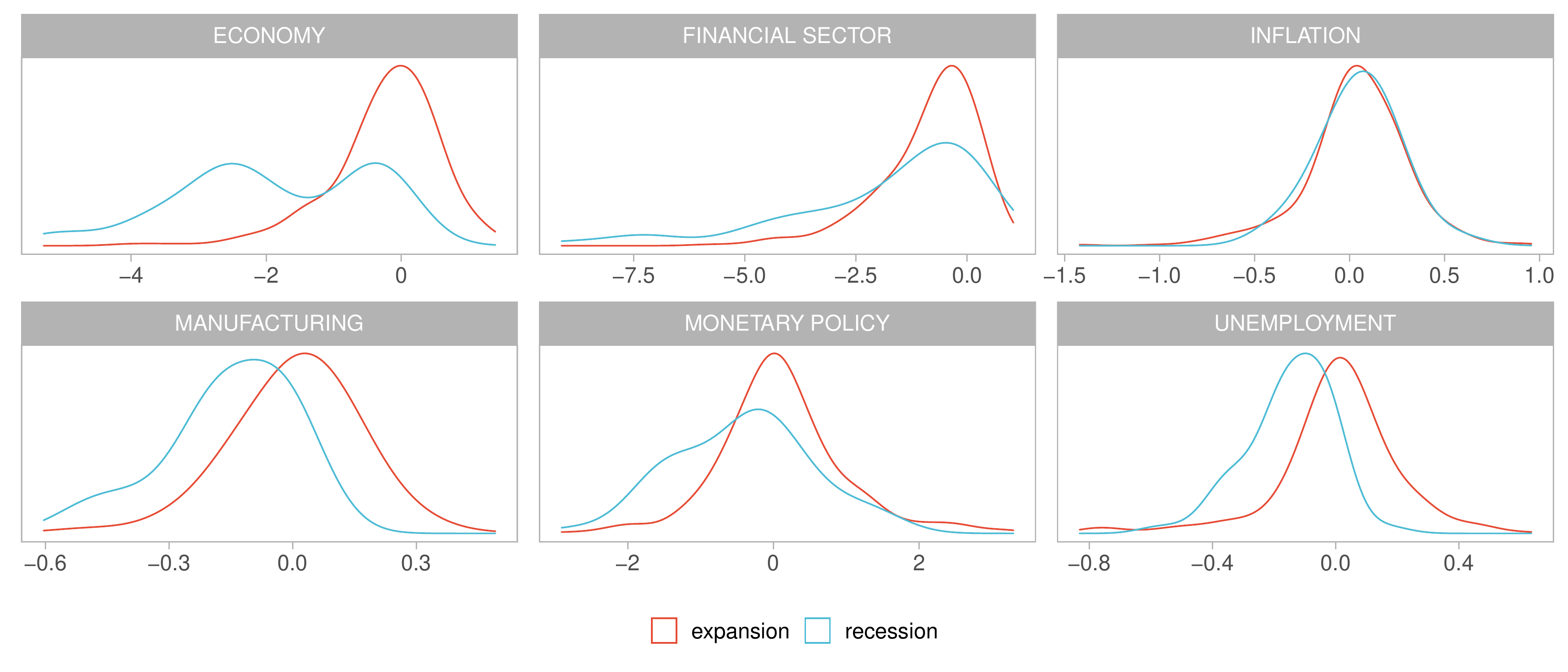}
 \caption{Kernel density of the economic sentiment measures for all verbal tenses separately for periods of expansion (red) and recession (green) as defined by the NBER business cycle committee. The density is calculated on the monthly series.}
 \label{fig:density_sentiment_all}
\end{figure}

%\bigskip\noindent
%% DENSITY
In Figure \ref{fig:density_sentiment_all} we calculate the smoothed distribution of the sentiment measures separately for expansions and recessions. For all measures, except unemployment, the density during recessions looks shifted to the left and shows a long left tail relative to the distribution during expansionary periods. Hence, the news sentiment about economic activity seems to capture the overall state of the economy and its transition between expansions and recessions. These graphs show also some interesting facts. One is the bimodality of the {\it economy} sentiment during recessions. 
Based on the time series graph in Figure \ref{fig:ts_sentiment_paprfunan}, the bimodality does not seem to arise because of recessions of different severity, as in the last three recessions the indicator reached a minimum smaller than -2. 
Rather, the bimodality seems to reflect different phases of a contraction, one state with sentiment slowly deteriorating followed by a jump to a state around the trough of the recession with rapidly deteriorating sentiment.
Another interesting finding is that sentiment about \textit{inflation} does not seem to vary significantly over the stages of the cycle. For {\it \textcolor{black}{manufacturing}} and {\it unemployment} the distribution of sentiment during recessions seems consistent with a shift of the mean of the distribution, while the variance appears to be similar in the two phases. 

\section{In-sample analysis\label{sec:in_sample}}
In this section we evaluate the statistical significance of the sentiment measures as predictors of economic activity on the full sample period that ranges from the beginning of 1980 to the end of 2019. 

\subsection{Significance of the economic sentiment measures\label{sec_var_sel}}

\begin{figure}[ht]
 \centering
\includegraphics[scale = 0.5]{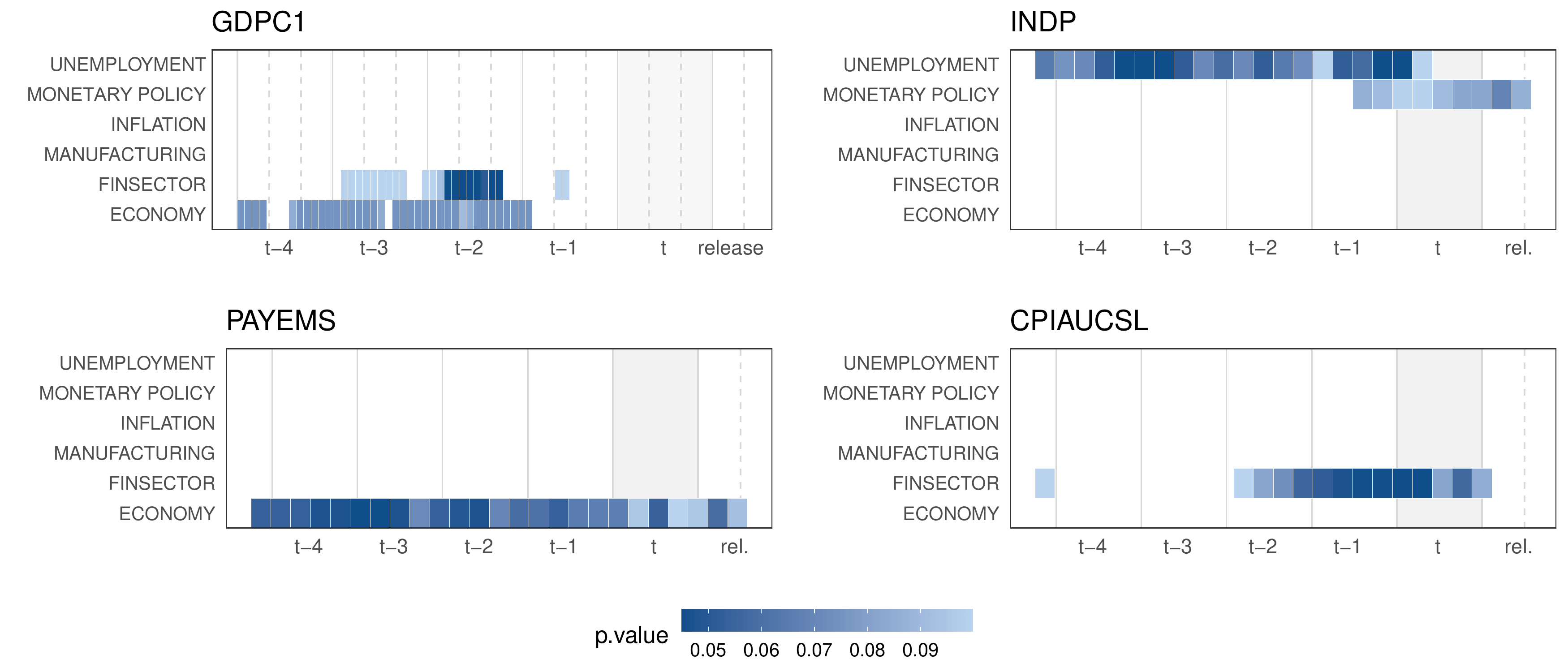}
 \caption{\small Statistical significance of the sentiment measures based on the double lasso penalized regression in Equation \eqref{eq:double_lasso_equation1} at each horizon $h$. 
 The statistic reported is the $p$-value of the sentiment coefficient $\eta_h$ corrected for multiple testing. 
 The colors depend on the $p$-value for those variables that are significant at least at 10\%: the darker the tile’s color, the smaller the $p$-value.
 The grey-shaded area corresponds to the reference period.}
 \label{fig:selection_insample}
\end{figure}

Figure \ref{fig:selection_insample} shows the statistical significance of the coefficient estimate associated to a sentiment measure, that is $\hat{\eta}_h$ in Equation \eqref{eq:double_lasso_equation1}, based on the double lasso procedure discussed in Section \ref{sec:sec1}. 
A tile at a given horizon $h$ indicates that the sentiment measure has a $p$-value smaller than 10\% and darker shades indicate smaller $p$-values. 
The fact that we are testing simultaneously many hypotheses might lead to spurious evidence of significance (see \citealp{james2021multiple} for a recent discussion). 
To account for the multiple testing nature of our exercise, we adjust the p-values associated to each sentiment reported in Figure~\ref{fig:selection_insample} for multiple testing based on the approach proposed by \cite{benjamini1995controlling} and \cite{benjamini2006adaptive}. 
The goal of the method is to control the False Discovery Rate that represents the expected rejection rate of a null hypothesis that is true. If we denote by $\hat p_j$ (for $j=1,\cdots, S$) the p-value of hypothesis $j$ sorted from smallest to largest, the adjusted p-value is given by $\hat p^{adj}_j = min\left(1, \hat p_j \frac{S}{j}\right)$. The power of the test can be improved if we assume that the number of null hypothesis that are true ($S_0$) is smaller relative to the number of hypotheses being tested ($S$). In this case the adjusted p-value is given by $\hat p^{adj}_j = min\left(1, \hat p_j \frac{S_0}{j}\right)$, where $S_0$ can be estimated following the two-step approach proposed by \cite{benjamini2006adaptive}. The first step of the procedure is to determine the number of rejected hypothesis ($R$) based on the \cite{benjamini1995controlling} adjusted p-values. $S_0$ is then set equal to $S-R$ and used to adjust the p-values as explained above\footnote{{
We correct for multiple testing across forecasting horizons by setting $S$ equal to the number of horizons (i.e., $69$ for quarterly GDP and $25$ for all monthly variables).  
This approach is consistent with the methodology followed in Section \ref{sec:out_of_sample} where we test the out-of-sample predictive ability jointly across horizons.}}. 

The top-left plot in Figure \ref{fig:selection_insample} reports the significance results for GDPC1.
We find that the most significant indicators are the \textit{financial sector} and \textit{economy} sentiments that are consistently selected at several horizons between quarter $t-4$ and $t-2$. This suggests that news sentiment at longer horizons is relevant to predict GDP growth, while at shorter horizons hard macroeconomic indicators become more relevant. 
%Interestingly, these measures are also significant at the beginning of the nowcasting quarter around the time GDP for quarter $t-1$ is released. This indicates that the macroeconomic release triggers news regarding the state of the economy that convey sentiment with predictive power beyond what is accounted for by the available hard data. For the other sentiment measures we find that there are some horizons in which they are statistically significant, although not persistently across horizons as in the case of \textit{financial sector} and \textit{economy}.
%% MONTHLY VARIABLES
%We find also interesting results for the monthly macroeconomic indicators. 
The results for the monthly indicators also show that the predictability is concentrated on a few sentiment measures. 
In particular, sentiment about \textit{unemployment} is significant at most horizons when forecasting INDP, while the {\it monetary policy} measure is significant at the nowcasting horizons. However, when predicting non-farm payroll employment (PAYEMS) we find that only the \textit{economy} indicator is a strongly significant predictor at all horizons. 
The \textit{financial sector} sentiment measure is also significant for CPI inflation starting from month $t-2$. 
%Interestingly, the selection exercise for CPIAUCSL shows that the proposed sentiment measures are mostly selected at short-term and nowcasting horizons, rather than at long forecasting horizons. This confirms that inflation is a very difficult series to forecast \citep{stock2007has}, although news sentiment seems to be helpful in nowcasting the series. 
Overall, our in-sample analysis suggests that predictability embedded in these news sentiment indicators is concentrated in one or two of the sentiment measures, depending on the variable being forecast. In addition, the horizons at which news sentiment matters seems also different across variables. While for the GPDC1 and INDP we find that the indicators are mostly relevant at long horizons, for CPIAUCSL the significance is mostly at the shortest horizons. Only in the case of PAYEMS we find that a sentiment indicators is significant at all horizons considered.

%% COEFFICIENTS TIME SERIES
%Figure \ref{fig:coeff_est} in the Appendix reports the ...

% \begin{figure}[ht]
% \centering
% \includegraphics[scale = 0.5]{Figures/double_lasso_localprojection_all.pdf}
% \label{fig:coeff_est}
% \caption{\small Estimate and 90\% confidence interval of the coefficient of the sentiment measures in Equation~(\ref{eq:double_lasso_equation1}) based on the double lasso penalized regression at weekly horizon $h$. The frequency is quarterly for GDPC1 and monthly for the remaining variables. }
% \label{fig:selection_insample}
% \end{figure}

\subsection{Quantile analysis}

\begin{figure}[b!]
 \centering
 \includegraphics[scale = 0.45]{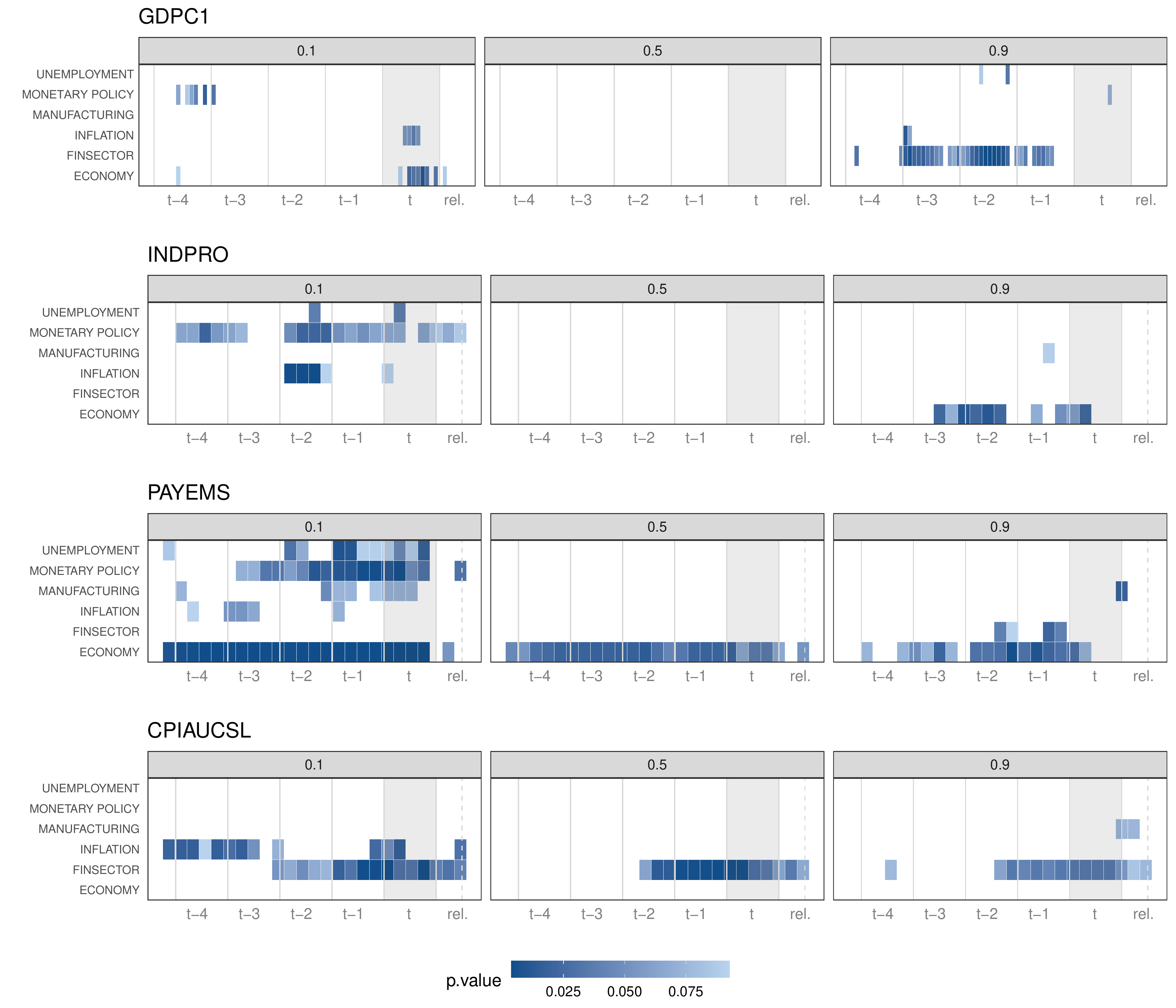}
 \caption{\small Significance of the sentiment measures based on the double lasso penalized quantile regression at each horizon $h$ for quantiles 0.1 (left column), 0.5 (center) and 0.9 (right). The colors depend on the $p$-value corrected for multiple testing for those variables that are significant at least at 10\%: the darker the tile's color, the smaller the $p$-value. The grey-shaded area corresponds to the reference period.}
 \label{fig:selection_q_insample}
\end{figure}

In Figure \ref{fig:selection_q_insample} we report the adjusted p-values smaller than 0.10 at each quantile level using the double lasso estimation procedure \citep{belloni2019valid}. 
%Consistently with the results in \cite{adrian2019vulnerable}, we find that the NFCI is a significant predictor at the lowest, and occasionally median, quantile level for GDPC1, INDPRO, and PAYEMS, while for CPIAUCSL it is selected at almost all horizons at the median and highest quantiles. On the contrary, we find that the CFNAI is mostly useful for nowcasting the output and employment variables, although for PAYEMS it is selected also at longer horizons. When predicting CPI inflation, the CFNAI is relevant at longer horizons and mostly on the left tail of the distribution. 
%% GDP
The evidence suggests that some sentiment measures are significant in forecasting GDPC1 at 0.1 and 0.9 quantiles. 
In particular, the \textit{economy} and \textit{inflation} sentiment indicators are significant on the left tail of the distribution at some nowcasting horizons that correspond to the release of the previous quarter figures. Instead, sentiment about the \textit{financial sector} is the most significant indicator at the highest quantile between quarter $t-3$ and $t-1$.
%Besides, only the \textit{economy} sentiment indicator is significant at the median for some horizons during quarter $t-2$.
Interestingly, no indicator is significant at the median, although we find that the {\it financial sector} and the {\it economy} measures are significant in predicting the mean. A possible explanation for this result is that the conditional mean is more influenced, relative to the median, by the extreme events that occurred during the financial crises of 2008-09. This is particularly true in the case of GDP given the relative short time series available.
%% MONTHLY
%% INDP
Similarly to the case of GDPC1, the results for INDPRO show that the significance of the sentiment measures is concentrated at low and high quantiles.
The sentiment about \textit{monetary policy} is selected at several horizons while the \textit{inflation} indicator is selected mostly during month $t-2$ on the left tail of the distribution. Instead, at the top quantile of the INDPRO distribution we find that the \textit{economy} sentiment measure is strongly significant at horizons between $t-3$ and $t$.
%% PAYEMS
The \textit{economy} indicator confirms to be a useful predictor for PAYEMS with strong significance across all quantiles and horizons considered. In addition, the \textit{unemployment}, \textit{manufacturing} and \textit{monetary policy} measures provide forecasting power at short horizons on the left tail of the distribution.
%% CPI
With respect to CPIAUCSL, the \textit{financial sector} indicator is statistically significant for horizons starting from $t-2$ to month $r$ at all quantiles confirming the results of Figure \ref{fig:selection_insample}. In addition, the sentiment about {\it inflation} is a relevant predictor at the lowest quantile for horizons from $t-3$, $t$, and $t-4$, and sporadically at shorter horizons.

%\begin{figure}[ht]
% \centering
% \includegraphics[scale = 0.7]{Figures/selection_quantile_insample.pdf}
% \caption{\small Sentiment measures selected by LASSO penalized quantile regression at each horizon $h$ for quantiles 0.1 (left column), 0.5 (center) and 0.9 (right). The variables selected are only those with a t-statistic larger than 1.64 in absolute terms and the color provides an indication of the sign of the coefficient, while the intensity of the color represents the size of the statistic.}
% \label{fig:selection_q_insample}
%\end{figure}

\section{Out-of-sample analysis\label{sec:out_of_sample}}

In this section we evaluate the robustness of the previous results to an out-of-sample test. We forecast the economic indicators from the beginning of 2002 until the end of 2019, a period that includes the Great Recession of 2008-2009. In our application, the benchmark for the relative accuracy test is the ARX specification while the alternative model augments the macroeconomic and financial information set with the sentiment indicators. 
 In addition, we also consider an \textit{Average} forecast obtained by averaging the ARXS forecasts and a \textit{LASSO} forecast that is obtained by lasso selection among a set of regressors that includes lagged values of the dependent variable, the three macroeconomic indicators, and the six sentiment indicators. We compare the forecast accuracy of the sentiment-augmented specifications at each individual horizon as well as jointly at all horizons using the average Superior Predictive Ability (aSPA) test proposed by \cite{quaedvlieg2021multi}\footnote{In the application we use a block length of 3 and 999 bootstrap replications as in \cite{quaedvlieg2021multi}.}. 
The evaluation of the point forecasts is performed using the square loss function that is defined as $f(e_{t,d-h}) = e_{t,d-h}^2$, where $f(\cdot)$ represents the loss function and $e_{t,d-h}$ is the error in forecasting the period $t$ release based on the information available on day $d-h$. Instead, the evaluation of the quantile forecasts is based on the check loss function defined as $f(e^{\tau}_{t,d-h}) = e^{\tau}_{t,d-h} \left(\tau - \mathbbm{1}_{e^{\tau}_{t,d-h} < 0}\right)$, where $\tau$ represents the quantile level and $e^{\tau}_{t,d-h} = Y_{t}^d - Q_{\tau,h}(Y_t^d)$. The choice of this loss function to evaluate quantiles is similar to \cite{giacomini2005evaluation} and is guided by the objective of aligning the loss used in estimation and in evaluation, as discussed in \cite{gneiting2011making}. In addition, we also consider the loss function $f(e^{\tau}_{t,d-h}) = (\tau - \mathbbm{1}_{e^{\tau}_{t,d-h} < 0})^2$ that evaluates the conditional coverage of the one-sided interval at level $\tau$. This loss function is relevant when the goal is to evaluate the coverage of an interval as in the applications to Growth-at-Risk at quantile levels 0.05 and 0.10 (see \citealp{adrian2019vulnerable} and \citealp{corradi2020conditional}).
%In addition to evaluate the accuracy of the model forecasts in the out-of-sample period, we also consider the fluctuation test proposed by \cite{giacomini2010forecast} which extends the statistical comparison to rolling windows over the out-of-sample period. 
Finally, the out-of-sample exercise is performed in real-time by providing the forecasting models with the vintage of macroeconomic information that was available at the time the forecast was made. We report results for the evaluation of the forecasts with respect to the second release, and results for the preliminary estimate are qualitatively similar.

\subsection{Performance of point forecasts}

Table \ref{tab:quad_point} reports the $p$-values of the \textit{aSPA} test for the null hypothesis that the forecasts from the alternative models do not outperform the benchmark forecasts at any horizons. The results for GDP indicate that the average forecasts have the lowest $p$-value at 0.143, although it is not statistically significant at 10\%. 
For INDP we find that sentiment about \textit{manufacturing}, \textit{monetary policy}, and \textit{unemployment} significantly improves the forecast accuracy relative to the ARX forecasts across all horizons. Instead, the sentiment about the {\it economy} delivers significantly more accurate forecasts for PAYEMS, while sentiment about the {\it financial sector} is useful when forecasting CPI inflation. 
The six sentiment indicators were designed to capture different aspects of economic activity, such as the real economy (\textit{manufacturing} and \textit{economy}), the price level (\textit{inflation}), the labor market (\textit{unemployment}), and the financial sector (\textit{financial sector} and \textit{monetary policy}). It is thus not surprising that only few of them are relevant to predict these different macroeconomic variables. 
Furthermore, the out-of-sample significance of the sentiment indicators confirms, to a large extent, the in-sample results provided in Figure \ref{fig:selection_insample}\footnote{In section D of the Appendix we perform a fluctuation analysis of the relative forecast performance based on the test proposed by \cite{giacomini2010forecast}. The findings suggest that the higher accuracy of news-based forecasts is episodic, in the sense that it emerges during specific periods while in other periods news do not seem to add predictive power relative to purely macro-based forecasts.}. 
As concerns the average forecasts, we find that they are significantly better for all the monthly variables while the \textit{LASSO} is significant only in the case of PAYEMS\footnote{
In the Appendix we include some additional results for the point forecasts. In particular, in Section \ref{appendinx_R2} we discuss the economic importance of the forecast gains by looking at the relative improvements in terms of goodness-of-fit and forecast error, while in Section \ref{sec_nsi} we compare the performance of the FiGAS-based sentiment indicators to the News Sentiment Index proposed by \cite{shapiro2020measuring}.}. 

\begin{table}[ht]
\centering
\setlength{\tabcolsep}{20pt}
\renewcommand{\arraystretch}{1.2}
\begin{tabular}[t]{lrrrr}
\hline\hline
Sentiment & GDPC1 & INDP & PAYEMS & CPIAUCSL\\ \hline
AVERAGE & 0.143 & {\bf0.006} & {\bf0.010} & {\bf0.020}\\
LASSO & 0.675 & 0.987 & {\bf0.051} & 0.987\\
ECONOMY & 0.646 & 0.831 & {\bf 0.008} & 0.801\\
FINANCIAL SECTOR & 0.923 & 0.980 & 0.180 & {\bf 0.030}\\
INFLATION & 0.988 & 0.972 & 0.921 & 0.791\\
MANUFACTURING & 0.990 & {\bf 0.035} & 0.177 & 0.365\\
MONETARY POLICY & 0.950 & {\bf 0.076} & 0.149 & 0.619\\
UNEMPLOYMENT & 0.920 & {\bf 0.098} & 0.983 & 0.801\\
\hline\hline
\end{tabular}
\caption{One-sided $p$-values of the aSPA multi-horizon test proposed by \cite{quaedvlieg2021multi}. The test compares the ARXS, Average, and LASSO forecasts against the ARX forecasts. }
 \label{tab:quad_point}
\end{table}

\subsection{Performance of quantile forecasts}

\begin{table}[ht!]
\setlength{\tabcolsep}{12pt}
\renewcommand{\arraystretch}{1.05}
\centering
\begin{tabular}[t]{llrrrr}
\hline\hline
Variable & Sentiment & $\tau=0.10$ & $\tau=0.50$ & $\tau=0.90$ & One-sided\\
\hline
GDP & AVERAGE & 0.968 & 0.237 & 0.115 & 0.956\\
 & LASSO & 0.533 & 0.968 & 0.351 & 0.995\\
 & ECONOMY & 0.990 & 0.913 & 0.977 & 0.988\\
 & FINANCIAL SECTOR & 0.988 & 0.989 & 0.848 & 0.986\\ 
 & INFLATION & 0.967 & 0.986 & 0.978 & 0.955\\
 & MANUFACTURING & 0.977 & 0.953 & 0.718 & 0.970\\
 & MONETARY POLICY & 0.965 & 0.904 & 0.944 & 0.961\\
 & UNEMPLOYMENT & 0.975 & 0.947 & 0.197 & 0.969\\ \hline
INDP & AVERAGE & {\bf0.019} & {\bf0.008} & {\bf0.033} & 0.589\\
 & LASSO & 0.993 & 0.435 & 0.992 & 0.997\\
 & ECONOMY & 0.987 & 0.982 & {\bf0.044} & 0.993\\
 & FINANCIAL SECTOR & 0.953 & 0.956 & 0.983 & {\bf0.008}\\ 
 & INFLATION & 0.891 & 0.954 & 0.948 & 0.960\\
 & MANUFACTURING & 0.902 & 0.389 & 0.985 & 0.989\\
 & MONETARY POLICY & 0.989 & 0.974 & 0.995 & 0.973\\
 & UNEMPLOYMENT & 0.631 & 0.643 & 0.958 & 0.165\\ \hline
PAYEMS & AVERAGE & {\bf0.003} & {\bf0.016} & {\bf0.022} & {\bf0.011}\\
 & LASSO & {\bf0.010} & 0.861 & 0.988 & 0.890\\
 & ECONOMY & {\bf0.016} & 0.144 & {\bf0.055} & 0.997\\
 & FINANCIAL SECTOR & {\bf0.077} & {\bf0.050} & 0.989 & {\bf0.023}\\ 
 & INFLATION & 0.987 & 0.972 & 0.985 & 0.943\\
 & MANUFACTURING & {\bf0.032} & 0.852 & 0.908 & 0.483\\
 & MONETARY POLICY & 0.172 & 0.337 & 0.993 & 0.990\\
 & UNEMPLOYMENT & 0.849 & 0.997 & 0.999 & {\bf0.009}\\ \hline
CPI & AVERAGE & {\bf0.019} & {\bf0.011} & 0.173 & 0.731\\
 & LASSO & {\bf0.051} & {\bf0.079} & 0.976 & 0.915\\
 & ECONOMY & 0.977 & 0.801 & 0.976 & {\bf0.095}\\
 & INFLATION & {\bf0.012} & {\bf0.010} & 0.341 & 0.933\\
 & FINANCIAL SECTOR & 0.889 & {\bf0.027} & 0.969 & 0.281\\
 & MANUFACTURING & 0.915 & 0.199 & 0.865 & 0.985\\
 & MONETARY POLICY & 0.605 & 0.923 & {\bf0.030} & {\bf0.050}\\
 & UNEMPLOYMENT & 0.941 & 0.925 & 0.973 & 0.894\\
\hline\hline
\end{tabular}
\caption{One-sided $p$-values of the \textit{aSPA} multi-horizon test proposed by \cite{quaedvlieg2021multi}. The test compares the ARXS, Average, and LASSO quantile forecasts against the ARX quantile forecasts for $\tau$ equal to 0.10, 0.50, and 0.9 (first three columns). We also evaluate the predictive ability on the one-sided interval from 0 to 0.10 (last column).}
 \label{tab:quad_quant}
\end{table}

Table \ref{tab:quad_quant} shows the results of the multi-horizon accuracy test applied to the quantile forecasts. The results for the individual quantile evaluation in the first three columns of the table indicate that the sentiment measures do not improve significantly when forecasting quarterly GDP, with only the average achieving $p$-values of 0.115 and 0.237 at the 0.5 and 0.9 quantiles, respectively. However, when predicting the monthly indicators we find that the average forecasts are significantly more accurate for all variables and for most quantile levels. The LASSO forecasting model outperforms the benchmark at the lowest quantile for PAYEMS, and at the median and lowest quantiles when forecasting CPI inflation. 
In terms of the ARXS specification, we find that the sentiment about the {\it economy} is useful to predict the tail quantiles of PAYEMS, while {\it financial sector} and {\it manufacturing} contribute to more accurate forecasts at the lowest quantile. Interestingly, we find that sentiment about {\it inflation} is a powerful predictor of future CPI inflation both at the lowest and median quantiles, while {\it financial sector} is only relevant at the median and {\it monetary policy} on the right-tail of the distribution.

The last column of Table~\ref{tab:quad_quant} shows the $p$-value of the multi-horizon test for the pairwise comparison of the conditional coverage of the left open interval at 10\%. We find that the {\it financial sector} sentiment delivers significantly more accurate interval forecasts for INDP and PAYEMS, while sentiment about the {\it economy} and {\it monetary policy} have more accurate coverage, relative to the benchmark, for CPI. Testing conditional coverage on the left tail of the distribution entails periods with large negative errors which, in our out-of-sample period, occurred to a large extent during the Great Recession of 2008-09. It is not thus surprising that sentiment deriving from news about the financial sector, the state of the economy and the conduct of monetary policy contributed to produce more accurate coverage on the left tail. 
Also, the table shows that in some cases the sentiment measures might be a significant predictor for a certain quantile but not for the interval. This result is consistent with the fact that different loss functions evaluate different aspects of the density forecast and thus might lead to contrasting conclusions regarding the significance of the sentiment measures.

\section{Conclusion}
\label{sec:sec3}

Macroeconomic forecasting has long relied on building increasingly complex econometric models to optimally use the information provided by statistical agencies. The recent availability of alternative data sets is leading the way for macroeconomists to create their own macroeconomic indicators to use along with the official statistics, thus enriching their information set. 
In this paper we provide an example of this new approach that uses over four billion words from newspaper articles over 40 years to create proxies for different aspects of economic sentiment. 
Our findings show the potential for using the sentiment measures in economic forecasting applications, providing  reliable measures to use together with official macroeconomic statistics.
An important advantage of using alternative data sets is that measures of economic activity can be constructed by the researcher at the daily frequency and in real-time, such as in the case of our sentiment indicators. In addition, the availability of the granular data opens the possibility for the researcher to investigate ways to produce more powerful indicators.
Our encouraging results indicate that sentiment extracted from text, and in particular news, is a promising route to follow, which could lead to significant improvements in macro-economic forecasting accuracy.
However, the application of text analysis in economics and finance is still in its infancy and more work is needed to understand its potential and relevance in different fields of economics and finance. 
\newpage
\spacingset{0.9} 
\bibliographystyle{chicago}
\bibliography{nowcasting.bib}

\begin{thebibliography}{}

\bibitem[\protect\citeauthoryear{Adrian, Boyarchenko, and Giannone}{Adrian
  et~al.}{2019}]{adrian2019vulnerable}
Adrian, T., N.~Boyarchenko, and D.~Giannone (2019).
\newblock Vulnerable growth.
\newblock {\em American Economic Review\/}~{\em 109\/}(4), 1263--89.

\bibitem[\protect\citeauthoryear{Adrian, Grinberg, Liang, and Malik}{Adrian
  et~al.}{2018}]{adrian2018term}
Adrian, T., F.~Grinberg, N.~Liang, and S.~Malik (2018).
\newblock The term structure of growth-at-risk.
\newblock {\em IMF working paper WP/18/180\/}.

\bibitem[\protect\citeauthoryear{Algaba, Ardia, Bluteau, Borms, and
  Boudt}{Algaba et~al.}{2020}]{algaba2020}
Algaba, A., D.~Ardia, K.~Bluteau, S.~Borms, and K.~Boudt (2020).
\newblock Econometrics meets sentiment: An overview of methodology and
  applications.
\newblock {\em Journal of Economic Surveys\/}~{\em 34\/}(3), 512--547.

\bibitem[\protect\citeauthoryear{Aprigliano, Ardizzi, Monteforte,
  et~al.}{Aprigliano et~al.}{2019}]{aprigliano2019using}
Aprigliano, V., G.~Ardizzi, L.~Monteforte, et~al. (2019).
\newblock Using the payment system data to forecast the economic activity.
\newblock {\em International Journal of Central Banking\/}~{\em 15\/}(4),
  55--80.

\bibitem[\protect\citeauthoryear{Aruoba, Diebold, and Scotti}{Aruoba
  et~al.}{2009}]{aruoba2009real}
Aruoba, S.~B., F.~X. Diebold, and C.~Scotti (2009).
\newblock Real-time measurement of business conditions.
\newblock {\em Journal of Business \& Economic Statistics\/}~{\em 27\/}(4),
  417--427.

\bibitem[\protect\citeauthoryear{Askitas and Zimmermann}{Askitas and
  Zimmermann}{2013}]{askitas2013nowcasting}
Askitas, N. and K.~F. Zimmermann (2013).
\newblock Nowcasting business cycles using toll data.
\newblock {\em Journal of Forecasting\/}~{\em 32\/}(4), 299--306.

\bibitem[\protect\citeauthoryear{Baccianella, Esuli, and
  Sebastiani}{Baccianella et~al.}{2010}]{baccianella2010sentiwordnet}
Baccianella, S., A.~Esuli, and F.~Sebastiani (2010).
\newblock Sentiwordnet 3.0: an enhanced lexical resource for sentiment analysis
  and opinion mining.
\newblock In {\em Lrec}, Volume~10, pp.\  2200--2204.

\bibitem[\protect\citeauthoryear{Baker, Bloom, and Davis}{Baker
  et~al.}{2016}]{baker2016measuring}
Baker, S.~R., N.~Bloom, and S.~J. Davis (2016).
\newblock Measuring economic policy uncertainty.
\newblock {\em Quarterly Journal of Economics\/}~{\em 131\/}(4), 1593--1636.

\bibitem[\protect\citeauthoryear{Baker, Bloom, Davis, and Kost}{Baker
  et~al.}{2019}]{baker2019policy}
Baker, S.~R., N.~Bloom, S.~J. Davis, and K.~J. Kost (2019).
\newblock Policy news and stock market volatility.
\newblock Technical report, National Bureau of Economic Research. Working paper
  25720.

\bibitem[\protect\citeauthoryear{Ba{\'n}bura, Giannone, Modugno, and
  Reichlin}{Ba{\'n}bura et~al.}{2013}]{banbura2013now}
Ba{\'n}bura, M., D.~Giannone, M.~Modugno, and L.~Reichlin (2013).
\newblock Now-casting and the real-time data flow.
\newblock In {\em Handbook of Economic Forecasting}, Volume~2, pp.\  195--237.
  Elsevier.

\bibitem[\protect\citeauthoryear{Belloni, Chernozhukov, and Hansen}{Belloni
  et~al.}{2014}]{belloni2014inference}
Belloni, A., V.~Chernozhukov, and C.~Hansen (2014).
\newblock Inference on treatment effects after selection among high-dimensional
  controls.
\newblock {\em The Review of Economic Studies\/}~{\em 81\/}(2), 608--650.

\bibitem[\protect\citeauthoryear{Belloni, Chernozhukov, and Kato}{Belloni
  et~al.}{2019}]{belloni2019valid}
Belloni, A., V.~Chernozhukov, and K.~Kato (2019).
\newblock Valid post-selection inference in high-dimensional approximately
  sparse quantile regression models.
\newblock {\em Journal of the American Statistical Association\/}~{\em
  114\/}(526), 749--758.

\bibitem[\protect\citeauthoryear{Benjamini and Hochberg}{Benjamini and
  Hochberg}{1995}]{benjamini1995controlling}
Benjamini, Y. and Y.~Hochberg (1995).
\newblock Controlling the false discovery rate: a practical and powerful
  approach to multiple testing.
\newblock {\em Journal of the Royal statistical society: series B\/}~{\em
  57\/}(1), 289--300.

\bibitem[\protect\citeauthoryear{Benjamini, Krieger, and Yekutieli}{Benjamini
  et~al.}{2006}]{benjamini2006adaptive}
Benjamini, Y., A.~M. Krieger, and D.~Yekutieli (2006).
\newblock Adaptive linear step-up procedures that control the false discovery
  rate.
\newblock {\em Biometrika\/}~{\em 93\/}(3), 491--507.

\bibitem[\protect\citeauthoryear{Blinder and Krueger}{Blinder and
  Krueger}{2004}]{blinder2004does}
Blinder, A.~S. and A.~B. Krueger (2004).
\newblock What does the public know about economic policy, and how does it know
  it?
\newblock Technical report, National Bureau of Economic Research, Working paper
  10787.

\bibitem[\protect\citeauthoryear{Bok, Caratelli, Giannone, Sbordone, and
  Tambalotti}{Bok et~al.}{2018}]{bok2018macroeconomic}
Bok, B., D.~Caratelli, D.~Giannone, A.~M. Sbordone, and A.~Tambalotti (2018).
\newblock Macroeconomic nowcasting and forecasting with big data.
\newblock {\em Annual Review of Economics\/}~{\em 10}, 615--643.

\bibitem[\protect\citeauthoryear{Brave}{Brave}{2009}]{brave2009chicago}
Brave, S. (2009).
\newblock The {C}hicago {F}ed {N}ational {A}ctivity {I}ndex and business
  cycles.
\newblock {\em Chicago Fed Letter\/}~{\em 268}, 1.

\bibitem[\protect\citeauthoryear{Brave and Butters}{Brave and
  Butters}{2014}]{brave2014nowcasting}
Brave, S.~A. and R.~Butters (2014).
\newblock Nowcasting using the {C}hicago {F}ed {N}ational {A}ctivity {I}ndex.
\newblock {\em Economic perspectives\/}~{\em 38\/}(1), 569--594.

\bibitem[\protect\citeauthoryear{Brave and Butters}{Brave and
  Butters}{2011}]{brave2011monitoring}
Brave, S.~A. and R.~A. Butters (2011).
\newblock Monitoring financial stability: A financial conditions index
  approach.
\newblock {\em Economic Perspectives\/}~{\em 35\/}(1), 22.

\bibitem[\protect\citeauthoryear{Brownlees and Souza}{Brownlees and
  Souza}{2021}]{brownlees2021backtesting}
Brownlees, C. and A.~B. Souza (2021).
\newblock Backtesting global growth-at-risk.
\newblock {\em Journal of Monetary Economics\/}~{\em 118}, 312--330.

\bibitem[\protect\citeauthoryear{Bybee, Kelly, Manela, and Xiu}{Bybee
  et~al.}{2019}]{bybee2019structure}
Bybee, L., B.~T. Kelly, A.~Manela, and D.~Xiu (2019).
\newblock The structure of economic news.
\newblock Technical report, National Bureau of Economic Research. Working paper
  26648.

\bibitem[\protect\citeauthoryear{Calomiris and Mamaysky}{Calomiris and
  Mamaysky}{2019}]{calomiris2019news}
Calomiris, C.~W. and H.~Mamaysky (2019).
\newblock How news and its context drive risk and returns around the world.
\newblock {\em Journal of Financial Economics\/}~{\em 133\/}(2), 299--336.

\bibitem[\protect\citeauthoryear{Cambria and Hussain}{Cambria and
  Hussain}{2015}]{sentic2015cambria}
Cambria, E. and A.~Hussain (2015).
\newblock {\em {Sentic Computing: A Common-Sense-Based Framework for
  Concept-Level Sentiment Analysis}}.
\newblock Springer.

\bibitem[\protect\citeauthoryear{Cambria, Li, Xing, Poria, and Kwok}{Cambria
  et~al.}{2020}]{Cambria2020105}
Cambria, E., Y.~Li, F.~Xing, S.~Poria, and K.~Kwok (2020).
\newblock {SenticNet 6: Ensemble Application of Symbolic and Subsymbolic AI for
  Sentiment Analysis}.
\newblock In {\em International Conference on Information and Knowledge
  Management (CIKM), Proceedings}, pp.\  105--114.

\bibitem[\protect\citeauthoryear{Choi and Varian}{Choi and
  Varian}{2012}]{choi2012predicting}
Choi, H. and H.~Varian (2012).
\newblock Predicting the present with {G}oogle {T}rends.
\newblock {\em Economic Record\/}~{\em 88}, 2--9.

\bibitem[\protect\citeauthoryear{Consoli, Barbaglia, and Manzan}{Consoli
  et~al.}{2022}]{consoli2021fine}
Consoli, S., L.~Barbaglia, and S.~Manzan (2022).
\newblock Fine-grained, aspect-based sentiment analysis on economic and
  financial lexicon.
\newblock {\em Knowledge-Based Systems (accepted for publication)\/}.
\newblock Pre-print available at SSRN 3766194.

\bibitem[\protect\citeauthoryear{Corradi, Fosten, and Gutknecht}{Corradi
  et~al.}{2020}]{corradi2020conditional}
Corradi, V., J.~Fosten, and D.~Gutknecht (2020).
\newblock Conditional quantile coverage: an application to growth-at-risk.
\newblock {\em Available at SSRN 3670575\/}.

\bibitem[\protect\citeauthoryear{Fornaro}{Fornaro}{2016}]{fornaro2016predicting}
Fornaro, P. (2016).
\newblock Predicting {F}innish economic activity using firm-level data.
\newblock {\em International Journal of Forecasting\/}~{\em 32\/}(1), 10--19.

\bibitem[\protect\citeauthoryear{Foroni, Marcellino, and Schumacher}{Foroni
  et~al.}{2015}]{foroni2015unrestricted}
Foroni, C., M.~Marcellino, and C.~Schumacher (2015).
\newblock Unrestricted mixed data sampling ({MIDAS}): {MIDAS} regressions with
  unrestricted lag polynomials.
\newblock {\em Journal of the Royal Statistical Society: Series A (Statistics
  in Society)\/}~{\em 178\/}(1), 57--82.

\bibitem[\protect\citeauthoryear{Galbraith and Tkacz}{Galbraith and
  Tkacz}{2018}]{galbraith2018nowcasting}
Galbraith, J.~W. and G.~Tkacz (2018).
\newblock Nowcasting with payments system data.
\newblock {\em International Journal of Forecasting\/}~{\em 34\/}(2), 366--376.

\bibitem[\protect\citeauthoryear{Gentzkow, Kelly, and Taddy}{Gentzkow
  et~al.}{2019}]{gentzkow2017text}
Gentzkow, M., B.~Kelly, and M.~Taddy (2019).
\newblock Text as data.
\newblock {\em Journal of Economic Literature\/}~{\em 57\/}(3), 535--74.

\bibitem[\protect\citeauthoryear{Gentzkow and Shapiro}{Gentzkow and
  Shapiro}{2010}]{gentzkow2010drives}
Gentzkow, M. and J.~M. Shapiro (2010).
\newblock What drives media slant? {Evidence from US} daily newspapers.
\newblock {\em Econometrica\/}~{\em 78\/}(1), 35--71.

\bibitem[\protect\citeauthoryear{Giacomini and Komunjer}{Giacomini and
  Komunjer}{2005}]{giacomini2005evaluation}
Giacomini, R. and I.~Komunjer (2005).
\newblock Evaluation and combination of conditional quantile forecasts.
\newblock {\em Journal of Business \& Economic Statistics\/}~{\em 23\/}(4),
  416--431.

\bibitem[\protect\citeauthoryear{Giacomini and Rossi}{Giacomini and
  Rossi}{2010}]{giacomini2010forecast}
Giacomini, R. and B.~Rossi (2010).
\newblock Forecast comparisons in unstable environments.
\newblock {\em Journal of Applied Econometrics\/}~{\em 25\/}(4), 595--620.

\bibitem[\protect\citeauthoryear{Gneiting}{Gneiting}{2011}]{gneiting2011making}
Gneiting, T. (2011).
\newblock Making and evaluating point forecasts.
\newblock {\em Journal of the American Statistical Association\/}~{\em
  106\/}(494), 746--762.

\bibitem[\protect\citeauthoryear{Hansen and McMahon}{Hansen and
  McMahon}{2016}]{hansen2016shocking}
Hansen, S. and M.~McMahon (2016).
\newblock Shocking language: Understanding the macroeconomic effects of central
  bank communication.
\newblock {\em Journal of International Economics\/}~{\em 99}, S114--S133.

\bibitem[\protect\citeauthoryear{Hansen, McMahon, and Prat}{Hansen
  et~al.}{2017}]{hansen2017transparency}
Hansen, S., M.~McMahon, and A.~Prat (2017).
\newblock Transparency and deliberation within the {FOMC}: a computational
  linguistics approach.
\newblock {\em Quarterly Journal of Economics\/}~{\em 133\/}(2), 801--870.

\bibitem[\protect\citeauthoryear{James, Witten, Hastie, and Tibshirani}{James
  et~al.}{2021}]{james2021multiple}
James, G., D.~Witten, T.~Hastie, and R.~Tibshirani (2021).
\newblock Multiple testing.
\newblock In {\em An Introduction to Statistical Learning}, pp.\  553--595.
  Springer.

\bibitem[\protect\citeauthoryear{Jord{\`a}}{Jord{\`a}}{2005}]{jorda2005estimation}
Jord{\`a}, {\`O}. (2005).
\newblock Estimation and inference of impulse responses by local projections.
\newblock {\em American economic review\/}~{\em 95\/}(1), 161--182.

\bibitem[\protect\citeauthoryear{Kalamara, Turrell, Kapetanios, Kapadia, and
  Redl}{Kalamara et~al.}{2018}]{kalamaramaking}
Kalamara, E., A.~Turrell, G.~Kapetanios, S.~Kapadia, and C.~Redl (2018).
\newblock Making text count for macroeconomics: What newspaper text can tell us
  about sentiment and uncertainty.
\newblock Technical report, Bank of England. Working paper 865.

\bibitem[\protect\citeauthoryear{Ke, Kelly, and Xiu}{Ke
  et~al.}{2019}]{ke2019predicting}
Ke, Z.~T., B.~T. Kelly, and D.~Xiu (2019).
\newblock Predicting returns with text data.
\newblock {\em Available in SSRN 3389884\/}.

\bibitem[\protect\citeauthoryear{Kelly, Manela, and Moreira}{Kelly
  et~al.}{2018}]{kelly2018text}
Kelly, B., A.~Manela, and A.~Moreira (2018).
\newblock Text selection.
\newblock {\em Available in SSRN 3491942\/}.

\bibitem[\protect\citeauthoryear{Koenker et~al.}{Koenker
  et~al.}{2011}]{koenker2011additive}
Koenker, R. et~al. (2011).
\newblock Additive models for quantile regression: Model selection and
  confidence bandaids.
\newblock {\em Brazilian Journal of Probability and Statistics\/}~{\em
  25\/}(3), 239--262.

\bibitem[\protect\citeauthoryear{Lamla and Maag}{Lamla and
  Maag}{2012}]{lamla2012role}
Lamla, M.~J. and T.~Maag (2012).
\newblock The role of media for inflation forecast disagreement of households
  and professional forecasters.
\newblock {\em Journal of Money, Credit and Banking\/}~{\em 44\/}(7),
  1325--1350.

\bibitem[\protect\citeauthoryear{Lewis, Mertens, and Stock}{Lewis
  et~al.}{2020}]{lewis2020us}
Lewis, D., K.~Mertens, and J.~H. Stock (2020).
\newblock {US} economic activity during the early weeks of the {SARS-Cov-2}
  outbreak.
\newblock Technical report, National Bureau of Economic Research.

\bibitem[\protect\citeauthoryear{Loughran and McDonald}{Loughran and
  McDonald}{2011}]{loughran2011liability}
Loughran, T. and B.~McDonald (2011).
\newblock {When is a liability not a liability? Textual analysis, dictionaries,
  and 10-Ks}.
\newblock {\em Journal of Finance\/}~{\em 66\/}(1), 35--65.

\bibitem[\protect\citeauthoryear{Manzan}{Manzan}{2015}]{manzan2015forecasting}
Manzan, S. (2015).
\newblock Forecasting the distribution of economic variables in a data-rich
  environment.
\newblock {\em Journal of Business \& Economic Statistics\/}~{\em 33\/}(1),
  144--164.

\bibitem[\protect\citeauthoryear{Manzan and Zerom}{Manzan and
  Zerom}{2013}]{manzan2013macroeconomic}
Manzan, S. and D.~Zerom (2013).
\newblock Are macroeconomic variables useful for forecasting the distribution
  of us inflation?
\newblock {\em International Journal of Forecasting\/}~{\em 29\/}(3), 469--478.

\bibitem[\protect\citeauthoryear{Marcellino and Schumacher}{Marcellino and
  Schumacher}{2010}]{marcellino2010factor}
Marcellino, M. and C.~Schumacher (2010).
\newblock Factor midas for nowcasting and forecasting with ragged-edge data: A
  model comparison for {German GDP}.
\newblock {\em Oxford Bulletin of Economics and Statistics\/}~{\em 72\/}(4),
  518--550.

\bibitem[\protect\citeauthoryear{Quaedvlieg}{Quaedvlieg}{2021}]{quaedvlieg2021multi}
Quaedvlieg, R. (2021).
\newblock Multi-horizon forecast comparison.
\newblock {\em Journal of Business \& Economic Statistics\/}~{\em 39\/}(1),
  40--53.

\bibitem[\protect\citeauthoryear{Shapiro, Sudhof, and Wilson}{Shapiro
  et~al.}{2020}]{shapiro2020measuring}
Shapiro, A.~H., M.~Sudhof, and D.~J. Wilson (2020).
\newblock Measuring news sentiment.
\newblock {\em Journal of Econometrics\/}~{\em (in press)}, 1--23.

\bibitem[\protect\citeauthoryear{Sharpe, Sinha, and Hollrah}{Sharpe
  et~al.}{2017}]{sharpe2017s}
Sharpe, S.~A., N.~R. Sinha, and C.~Hollrah (2017).
\newblock What's the story? {A} new perspective on the value of economic
  forecasts.
\newblock {\em FEDS Working Paper, Finance and Economics Discussion Series
  2017-107\/}.

\bibitem[\protect\citeauthoryear{Stock and Watson}{Stock and
  Watson}{1999}]{stock1999forecasting}
Stock, J.~H. and M.~W. Watson (1999).
\newblock Forecasting inflation.
\newblock {\em Journal of Monetary Economics\/}~{\em 44\/}(2), 293--335.

\bibitem[\protect\citeauthoryear{Stock and Watson}{Stock and
  Watson}{2016}]{stock2016dynamic}
Stock, J.~H. and M.~W. Watson (2016).
\newblock Dynamic factor models, factor-augmented vector autoregressions, and
  structural vector autoregressions in macroeconomics.
\newblock In {\em Handbook of Macroeconomics}, Volume~2, pp.\  415--525.
  Elsevier.

\bibitem[\protect\citeauthoryear{Tetlock}{Tetlock}{2007}]{tetlock2007}
Tetlock, P.~C. (2007).
\newblock Giving content to investor sentiment: The role of media in the stock
  market.
\newblock {\em Journal of Finance\/}~{\em 62\/}(3), 1139--1168.

\bibitem[\protect\citeauthoryear{Thorsrud}{Thorsrud}{2016}]{thorsrud2016nowcasting}
Thorsrud, L.~A. (2016).
\newblock Nowcasting using news topics. big data versus big bank.
\newblock {\em Norges Bank Working Paper 20/2016\/}.

\bibitem[\protect\citeauthoryear{Thorsrud}{Thorsrud}{2020}]{thorsrud2020words}
Thorsrud, L.~A. (2020).
\newblock Words are the new numbers: A newsy coincident index of the business
  cycle.
\newblock {\em Journal of Business \& Economic Statistics\/}~{\em 38\/}(2),
  393--409.

\bibitem[\protect\citeauthoryear{Tibshirani}{Tibshirani}{1996}]{tibshirani1996regression}
Tibshirani, R. (1996).
\newblock Regression shrinkage and selection via the lasso.
\newblock {\em Journal of the Royal Statistical Society: Series B
  (Methodological)\/}~{\em 58\/}(1), 267--288.

\bibitem[\protect\citeauthoryear{Xing, Cambria, and Welsch}{Xing
  et~al.}{2018}]{Xing201849}
Xing, F., E.~Cambria, and R.~Welsch (2018).
\newblock Natural language based financial forecasting{: A} survey.
\newblock {\em Artificial Intelligence Review\/}~{\em 50\/}(1), 49--73.

\end{thebibliography}

\appendix

\newpage
\spacingset{1.2} 

\if0\blind{
\begin{center}{\LARGE\bf Appendix for ``Forecasting with Economic News''}\\
\mbox{}\\
{Luca Barbaglia$^a$, Sergio Consoli$^a$, Sebastiano Manzan$^b$}\\
{$^a$ Joint Research Centre, European Commission}\\
{$^b$ Zicklin School of Business, Baruch College}\\
\end{center}
\vspace{2cm}
}\fi
\if1\blind{
\begin{center}
{\LARGE\bf Appendix for ``Forecasting with Economic News''}
\end{center}
\vspace{2cm}
}\fi

\section{Description of the FiGAS algorithm
\label{Appendix_FiGAS}}

In this section we provide additional details regarding the linguistic rules that are used in the NLP workflow discussed in Section \ref{sec:sec1} of the main paper. 
For a complete presentation of the FiGAS algorithm and a comparison with alternative sentiment analysis methods, we refer to \cite{consoli2021fine}.
Our analysis is based on the \textit{en\_core\_web\_lg} language model available in the Python module \textit{spaCy}\footnote{\url{https://spacy.io/api/annotation}.}. The package is used to assign to each word a category for the following properties: 
\begin{itemize}
 \item POS : part-of-speech; 
 \item TAG : tag of the part-of-speech;
 \item DEP : dependency parsing.%\footnote{See \url{https://universaldependencies.org/u/dep/} for more details on the dependency parsing employed spaCy.}.
\end{itemize}
Tables \ref{tab_Spacy_POS} and \ref{tab_Spacy_Dependency} provides the labels used by spaCy for part-of-speech tagging and dependency parsing, respectively.
In the remainder of this section, we detail the steps of our NLP workflow, namely the tense detection, location filtering, semantic rules and score propagation based on the proposed fine-grained economic dictionary. 

%If you want to create an example of Spacy output, use \url{https://explosion.ai/demos/displacy}. 

\begin{table}[ht]
\medskip
\caption{Part-Of-Speech (POS) tagging. \label{tab_Spacy_POS}}
\centering
\medskip
\begin{tabular}{lll}
TAG & POS & DESCRIPTION \\
\hline
% -LRB- & PUNCT & left round bracket \\
% -RRB- & PUNCT & right round bracket \\
% , & PUNCT & punctuation mark, comma \\
% : & PUNCT & punctuation mark, colon or ellipsis \\
% . & PUNCT & punctuation mark, sentence closer \\
% '' & PUNCT & closing quotation mark \\
% "" & PUNCT & closing quotation mark \\
% `` & PUNCT & opening quotation mark \\
% \# & SYM & symbol, number sign \\
% \$ & SYM & symbol, currency \\
% ADD & X & email \\
% AFX & ADJ & affix \\
% BES & VERB & auxiliary â€œbeâ€ \\
CC & CONJ & conjunction, coordinating \\
% CD & NUM & cardinal number \\
% DT & DET & determiner \\
% EX & ADV & existential there \\
% FW & X & foreign word \\
% GW & X & additional word in multi-word expression \\
% HVS & VERB & forms of â€œhaveâ€ \\
% HYPH & PUNCT & punctuation mark, hyphen \\
IN & ADP & conjunction, subordinating or preposition \\
 JJ & ADJ & adjective \\
 JJR & ADJ & adjective, comparative \\
 JJS & ADJ & adjective, superlative \\
% LS & PUNCT & list item marker \\
MD & VERB & verb, modal auxiliary \\
% NFP & PUNCT & superfluous punctuation \\
% NIL & & missing tag \\
NN & NOUN & noun, singular or mass \\
NNP & PROPN & noun, proper singular \\
NNPS & PROPN & noun, proper plural \\
NNS & NOUN & noun, plural \\
% PDT & ADJ & predeterminer \\
% POS & PART & possessive ending \\
% PRP & PRON & pronoun, personal \\
% PRP\$ & ADJ & pronoun, possessive \\
% RB & ADV & adverb \\
RBR & ADV & adverb, comparative \\
RBS & ADV & adverb, superlative \\
% RP & PART & adverb, particle \\
% \_SP & SPACE & space \\
% SYM & SYM & symbol \\
% TO & PART & infinitival â€œtoâ€ \\
% UH & INTJ & interjection \\
VB & VERB & verb, base form \\
VBD & VERB & verb, past tense \\
VBG & VERB & verb, gerund or present participle \\
VBN & VERB & verb, past participle \\
VBP & VERB & verb, non-3rd person singular present \\
VBZ & VERB & verb, 3rd person singular present \\
% WDT & ADJ & wh-determiner \\
% WP & NOUN & wh-pronoun, personal \\
% WP\$ & ADJ & wh-pronoun, possessive \\
% WRB & ADV & wh-adverb \\
% XX & X & unknown 
\end{tabular}
\end{table}

\begin{table}[t!]
\medskip
\caption{spaCy dependency parsing. \label{tab_Spacy_Dependency}}
\centering
\medskip
\begin{tabular}{ll}
 DEP & DESCRIPTION \\
\hline
acl & clausal modifier of noun (adjectival clause) \\
% acomp & adjectival complement \\
advcl & adverbial clause modifier \\
advmod & adverbial modifier \\
% agent & agent \\
amod & adjectival modifier \\
% appos & appositional modifier \\
attr & attribute \\
% aux & auxiliary \\
% auxpass & auxiliary (passive) \\
% case & case marking \\
% cc & coordinating conjunction \\
% ccomp & clausal complement \\
% compound & compound \\
% conj & conjunct \\
% cop & copula \\
% csubj & clausal subject \\
% csubjpass & clausal subject (passive) \\
% dative & dative \\
% dep & unclassified dependent \\
% det & determiner \\
dobj & direct object \\
% expl & expletive \\
% intj & interjection \\
% mark & marker \\
% meta & meta modifier \\
neg & negation modifier \\
% nn & noun compound modifier \\
% nounmod & modifier of nominal \\
% npmod & noun phrase as adverbial modifier \\
% nsubj & nominal subject \\
% nsubjpass & nominal subject (passive) \\
% nummod & numeric modifier \\
oprd & object predicate \\
% obj & object \\
% obl & oblique nominal \\
% parataxis & parataxis \\
pcomp & complement of preposition \\
pobj & object of preposition \\
% poss & possession modifier \\
% preconj & pre-correlative conjunction \\
prep & prepositional modifier \\
% prt & particle \\
% punct & punctuation \\
% quantmod & modifier of quantifier \\
% relcl & relative clause modifier \\
% root & root \\
 xcomp & open clausal complement \\
% vbn & verb \\
% vbp & verb \\
% vbz & verb \\
% wdt & adj \\
% wp & noun \\
% wp\$ & adj \\
% wrb & adv \\
% XX & X 
\end{tabular}
\end{table}

\subsection{Tense}

We rely on the spaCy tense detection procedure to associate a verb ({\tt POS == VERB}) to one of the following tenses:
\begin{enumerate}
 \item[a.] past (i.e., {\tt TAG == VBD} or {\tt TAG == VBN});
 \item[b.] present (i.e., {\tt TAG == VBP} or {\tt TAG == VBZ} or {\tt TAG == VBG});
 \item[c.] future (i.e., {\tt TAG == MD}).
\end{enumerate}
If no tense is detected, spaCy assigns the ``NaN" value.
For each text chunk, we compute a score of each of the tenses above: the score is based on the number of verbs detected with that tense, weighted by the distance of the verb to the ToI (the closer the verb to the ToI, the larger the score). We then assign the tense with the highest score to the whole chunk.

\subsection{Location}
The location detection builds on the ability of the spaCy library to perform named-entity recognition. For our location task, we use the following entities:
\begin{itemize}
 \item countries, cities, states ({\tt GPE});
 \item nationalities or religious or political group ({\tt NORP});
 \item non-GPE locations, mountain ranges, bodies of water ({\tt LOC});
 \item companies, agencies, institutions, etc. ({\tt ORG}).
\end{itemize}
\noindent
We search for the desired location among the above entities with the following procedure. We first label each article with its most frequent location among the above recognized named-entities. Then, we look for the most frequent location in each chunk: if there are no entities, we then assign the article location also to the chunk.
FiGAS continues considering the specific chunk only if the assigned location is one of the locations of interest to the user, which in our case corresponds to terms referring to the United States\footnote{The \textit{locations and organizations} used for restricting the searches to US articles only are: America, United States, Columbia, land of liberty, new world, U.S., U.S.A., USA, US, land of opportunity, the states, Fed, Federal Reserve Board, Federal Reserve, Census Bureau, Bureau of Economic Analysis, Treasury Department, Department of Commerce, Bureau of Labor Statistics, Bureau of Labour, Department of Labor, Open Market Committee, BEA, Bureau of Economic Analysis, BIS, Bureau of Statistics, Board of Governors, Congressional Budget Office, CBO, Internal Revenue Service, IRS.}.

\subsection{Rules}

After the detection of the location and verbal tense, the FiGAS analysis continues with the evaluation of the semantic rules and the calculation of the sentiment value. 
In particular, the algorithm selects the text that satisfies the rules discussed below and assigns a sentiment value only to the tokens related to the ToI. Furthermore, if the chunk contains a negation or a term with a negative connotation ({\tt DEP == neg}), the final sentiment polarity of the whole chunk is reversed (\textit{negation handling}).
Below is the list of rules that we use to parse dependence:

\begin{enumerate}
\item{\it ToI associated to an adjectival modifier \label{DNA_rule_amod}}\\
The ToI is associated to a lemma with an adjectival modifier dependency ({\tt DEP == amod}) such as:
\begin{enumerate}
 \item[a.] an adjective ({\tt POS == ADJ}) in the form of (i) standard adjective ({\tt TAG == JJ}), (ii) comparative adjective ({\tt TAG == JJR}), or (iii) superlative adjective ({\tt TAG == JJS})
 \item[b.] a verb ({\tt POS == VERB}). 
\end{enumerate}
\noindent
\textit{Example (a.)} We observe a {\it stronger} {\sf industrial production} in the US than in other countries.\\
\textit{Example (b.)} We observe a {\it rising} {\sf industrial production} in the US.
\item {\it ToI associated to a verb in the form of an open clausal complement or an adverbial clause modifier \label{DNA_rule_xcomp_advcl}}\\
The ToI is related to a verb that is dependent on:
\begin{itemize}
 \item[a.] an open clausal complement (i.e., adjective is a predicative or clausal complement without its own subject; {\tt DEP == xcomp}); or
 \item[b.] an adverbial clause modifier (i.e., an adverbial clause modifier is a clause which modifies a verb or other predicate -adjective, etc.-, as a modifier not as a core complement; {\tt DEP == advcl}).
\end{itemize}
\noindent
\textit{Example (a.)} We expect {\sf output} to {\it grow}. \\
\textit{Example (b.)} {\sf Output} dipped, {\it leaving} it lower than last month.
\item{\it ToI associated to a verb followed by an adjective in the form of an adjectival complement \label{DNA_rule_acomp}}\\
The ToI is associated to a verb that is followed by an adjectival complement (i.e., a phrase that modifies an adjective and offers more information about it; {\tt DEP == acomp}) such as: 
\begin{itemize}
 \item[a.] standard adjective ({\tt TAG == JJ});
 \item[b.] comparative adjective ({\tt TAG == JJR});
 \item[c.] superlative adjective ({\tt TAG == JJS}).
\end{itemize}
\noindent
\textit{Example (a.)} The {\sf economic outlook} looks {\it good}. 
\item {\it ToI associated to a verb followed by an adjective in the form of an object predicate \label{DNA_rule_oprd}}\\
The ToI is associated to a verb followed by an object predicate (i.e. an adjective, noun phrase, or prepositional phrase that qualifies, describes, or renames the object that appears before it; {\tt DEP == oprd}) such as: 
\begin{itemize}
 \item[a.] standard adjective ({\tt TAG == JJ});
 \item[b.] comparative adjective ({\tt TAG == JJR});
 \item[c.] superlative adjective ({\tt TAG == JJS}).
\end{itemize}
\noindent
\textit{Example (b.)} The {\sf FED} kept the rates {\it lower} than expected.
\item{\it ToI associated to a verb followed by an adverbial modifier \label{DNA_rule_advmod}}\\
The ToI is associated to a verb followed by an adverbial modifier (i.e. a non-clausal adverb or adverbial phrase that serves to modify a predicate or a modifier word; {\tt DEP == advmod}) defined as:
\begin{itemize}
 \item[a.] a comparative adverb ({\tt TAG == RBR}); 
 \item[b.] a superlative adverb ({\tt TAG == RBS}).
\end{itemize}
\noindent
\textit{Example (a.)} UK retail {\sf sales} fared {\it better} than expected.

\item{\it ToI associated to a verb followed by a noun in the form of direct object or attribute \label{DNA_rule_dobj_attr}}\\
ToI associated to a verb followed by a noun in the form of:
\begin{itemize}
 \item[a.] a direct object (noun phrase which is the (accusative) object of the verb. {\tt DEP == dobj});
 \item[b.] an attribute (it qualifies a noun about its characteristics. {\tt DEP == attr}).
\end{itemize} 
\noindent
\textit{Example (a.)} The {\sf economy} suffered a {\it slowdown}. \\
\textit{Example (b.)} Green {\sf bonds} have been a {\it good} investment.
\item {\it ToI associated to a verb followed by a prepositional modifier}\\
ToI associated to a verb followed by a prepositional modifier (i.e. words, phrases, and clauses that modify or describe a prepositional phrase. {\tt DEP == prep}) and by a: 
\begin{itemize}
 \item[a.] a noun in the form of an object of preposition (i.e., a noun that follows a preposition and completes its meaning. {\tt DEP == pobj});
 \item[b.] a verb in the form of a complement of preposition (i.e., a clause that directly follows the preposition and completes the meaning of the prepositional phrase. {\tt DEP == pcomp}).
\end{itemize}
\noindent
\textit{Example (a.)} {\sf Markets} are said to be in {\it robust} form. \\
\textit{Example (b.)} Forecasts are bad, with a falling output.
\item {\it ToI associated to an adjectival clause}\\
ToI associated to an adjectival clause (i.e., a finite and non-finite clause that modifies a nominal. {\tt DEP == acl}).
\medskip
\noindent
\begin{itemize}
 \item \textit{Example} There are many online tools providing investing platforms.
\end{itemize}
\end{enumerate}

\subsection{Sentiment polarity propagation}\label{sec:propagation}

The following step is to compute the sentiment of the text. In case there are terms wiht non-zero sentiment (excluding the ToI), we proceed by:
\begin{enumerate}
 \item[i.] summing the sentiment of the term closer to the ToI with the sentiment of the remaining terms weighted by the tone of the closest term;
 \item[ii.] the sentiment value of the ToI is not included in the computation and it is only used to invert the sign of the resulting sentiment in case the ToI has negative sentiment.
\end{enumerate}

\subsection{Fine-grained economic and financial dictionary\label{Appedix_DICTIONARY}}
%% DICTIONARY
%The sentiment value that is assigned to each term in the FiGAS analysis derives from a custom dictionary that we created on purpose for the task.
Several applications in economics and finance compute sentiment measures based on the dictionary proposed by \cite{loughran2011liability} (LMD). The LMD dictionary includes approximately 4,000 words extracted from 10K filing forms that are classified in one of the following categories: negative, positive, uncertainty, litigious, strong, weak modals, and constraining. 
A limitation of the LMD dictionary is that it does not provide a continuous measure (e.g., between $\pm$ 1) of the strength of the sentiment carried by a word. 
%In addition, in the LMD dictionary negative words represent more than half of the total. 
%%%% SEBA: not clear what this means, the other half could be positive, 
%%%% so it is not clear the message we are trying to convey here ...
%%%% what is wrong with "half" negative? and what is the other "half"?

%\bigskip\noindent
To overcome this limitation, we enriched the LMD dictionary by extending the list of words and by assigning a numerical value to the sentiment. More specifically, we removed a small fraction of the 10K-specific terms not having polarity (e.g., juris, obligor, pledgor), and added to the dictionary a list of words that were frequently occurring in the news text presented in Section \ref{Sec_data} of the main paper. Next, we annotated each word in the expanded list by employing a total of 15 human annotators.
In particular, we collected 5 annotations from field experts\footnote{These experts are the authors of this paper and two economist colleagues at the Joint Research Centre.}, and additional 10 from US based professional annotators. 
The annotation task consisted of assigning a score to each term based on the annotator's perception of its negative, neutral or positive tone in the context of an article that discusses economic and financial issues. 
We required annotators to provide a score in the interval $[-1,1]$, with precision equal to one decimal. We also reminded annotators that a negative values $[-1,0)$ would indicate that the word is, for the average person, used with a negative connotation in an economic context, whereas positive scores $(0,1]$ would characterize words that are commonly used to indicate a favourable economic situation. 
The annotator was given the option to assign the value 0 if he/she believed that the word was neither negative nor positive. 
The magnitude of the score represents the sentiment intensity of the word as perceived by the annotator: for instance, a score close to 1 indicates an extremely positive economic situation, while a small positive value would be associated with a moderately favourable outcome. 
In case the annotator was doubtful about the tonality of a word, we provided an example of the use of the word in a sentence containing that specific term. These examples were taken at random from the data set presented in Section \ref{Sec_data}.

\begin{figure}[b!]
 \centering
 \includegraphics[scale = 0.55]{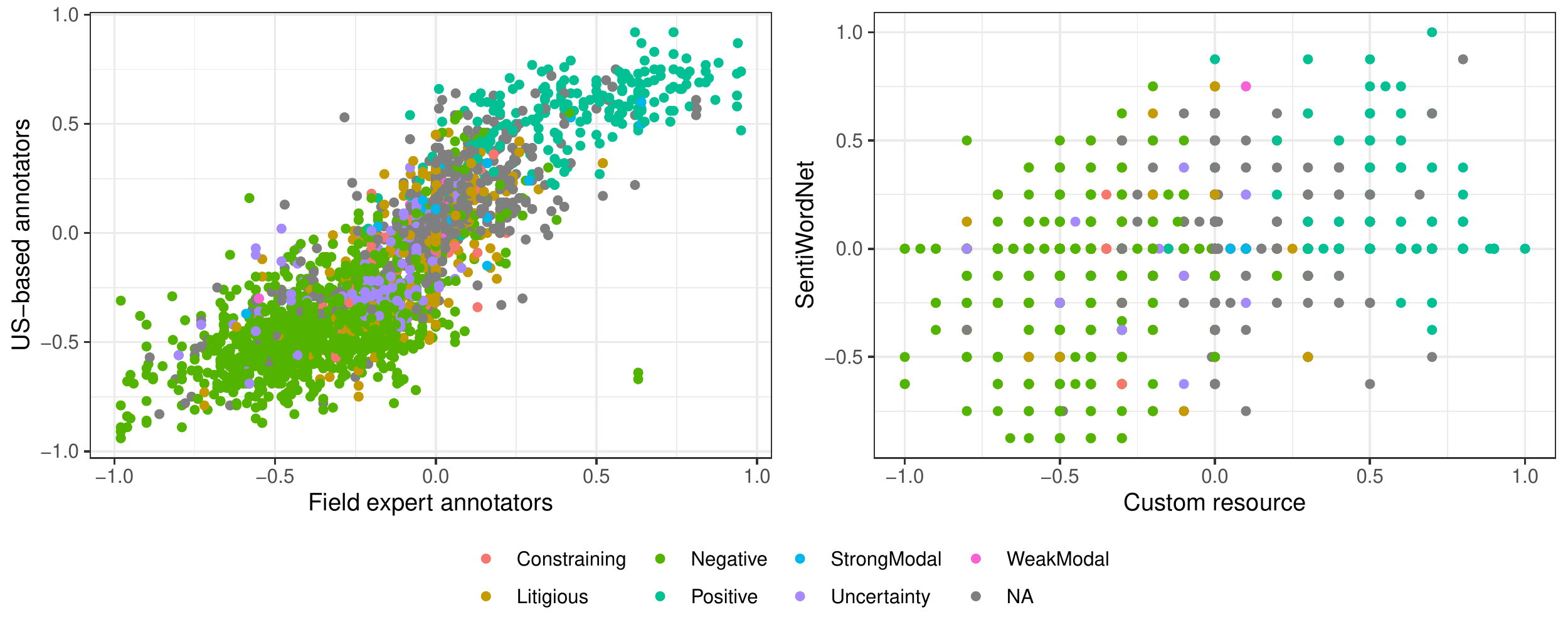}
 \caption{Mean scores US-based annotators versus field experts annotations (left panel), and SentiWordNet scores versus median score of the proposed resource (right panel). Colors indicates the categories identified by \cite{loughran2011liability}.}
 \label{fig:AWS_CAS2}
\end{figure}

%\bigskip\noindent
The left panel of Figure \ref{fig:AWS_CAS2} compares the mean scores assigned by the US-based and field expert annotators. 
In general, there seems to be an overall agreement between the two annotator groups, as demonstrated by the positive correlation between the average scores.
Moreover, the positivity or negativity of the resulting polarity scores have been verified with the categories of the \cite{loughran2011liability} dictionary, which are identified by colors in the Figure. In the right panel of Figure \ref{fig:AWS_CAS2} we compare the median score across the 15 annotators with the score in the SentiWordNet dictionary proposed by \cite{baccianella2010sentiwordnet}. Overall, there seems to be a weak positive correlation between the two resources, suggesting that often there is disagreement between the general purpose and the field-specific dictionaries. The resulting dictionary contains 3,181 terms with a score in the interval $\pm$1. We plan to further extend the dictionary to represent a comprehensive resource to analyze text in economics and finance.

%\bigskip\noindent
We now discuss an example that illustrates the FiGAS approach to sentiment analysis. Let's consider the following sentence: \textit{the US manufacturing sector has carried the brunt of the global economic slowdown over the past few months}. The FiGAS methodology would proceed in the following steps to compute sentiment for, e.g., the ToI {\it manufacturing sector}:
\begin{itemize}
\item The ToI {\sf manufacturing sector} is detected by looping on the part-of-speech tags; 
\item The ToI depends on {\it to carry} ({\tt POS == VERB}), which is neutral (i.e., sentiment equal to 0) based on the sentiment dictionary; 
\item The dependence analysis further shows that the \texttt{VERB} {\it to carry} is linked to the \texttt{DOBJ} (\textit{direct object}) \textit{brunt} which is a \texttt{NOUN} and has negative sentiment equal to -0.125 according to the dictionary;
\item Polarity is propagated as follows \textit{brunt} $\rightarrow$ \textit{to carry} $\rightarrow$ \textsf{manufacturing sector} and the final sentiment score of the ToI \textit{manufacturing sector} in this sentence is -0.125
\end{itemize}

\section{Verbal tense\label{sec_verbal}}

%% INTRO
As discussed in the previous section and in the main paper, we designed the FiGAS methodology to detect the tense of the verb that is related to the ToI, if any. The objective of this feature is to add a temporal dimension to the sentiment measures that could be useful when forecasting macroeconomic variables. More specifically, we calculate sentiment separately for the chunks of text in which we find that the verbal form is either {\it past}, {\it future}, or {\it present}. Instead, we denote by {\it NaN} the cases when the ToI is not dependent on a verb.
Combining the six sentiment measures with the four verbal tenses provides a total of 24 combinations of sentiment-tense measures. While in the paper we provide the results for the aggregate (by tense) sentiment indicators, the aim of this section is to discuss the potential benefits of separating the verbal tenses. We evaluate the statistical significance of the sentiment-tense measures based on the double lasso methodology presented in Section \ref{sec:sec1} of the paper.

%% MEAN

\begin{figure}[ht!]
 \centering
 \includegraphics[scale=0.5]{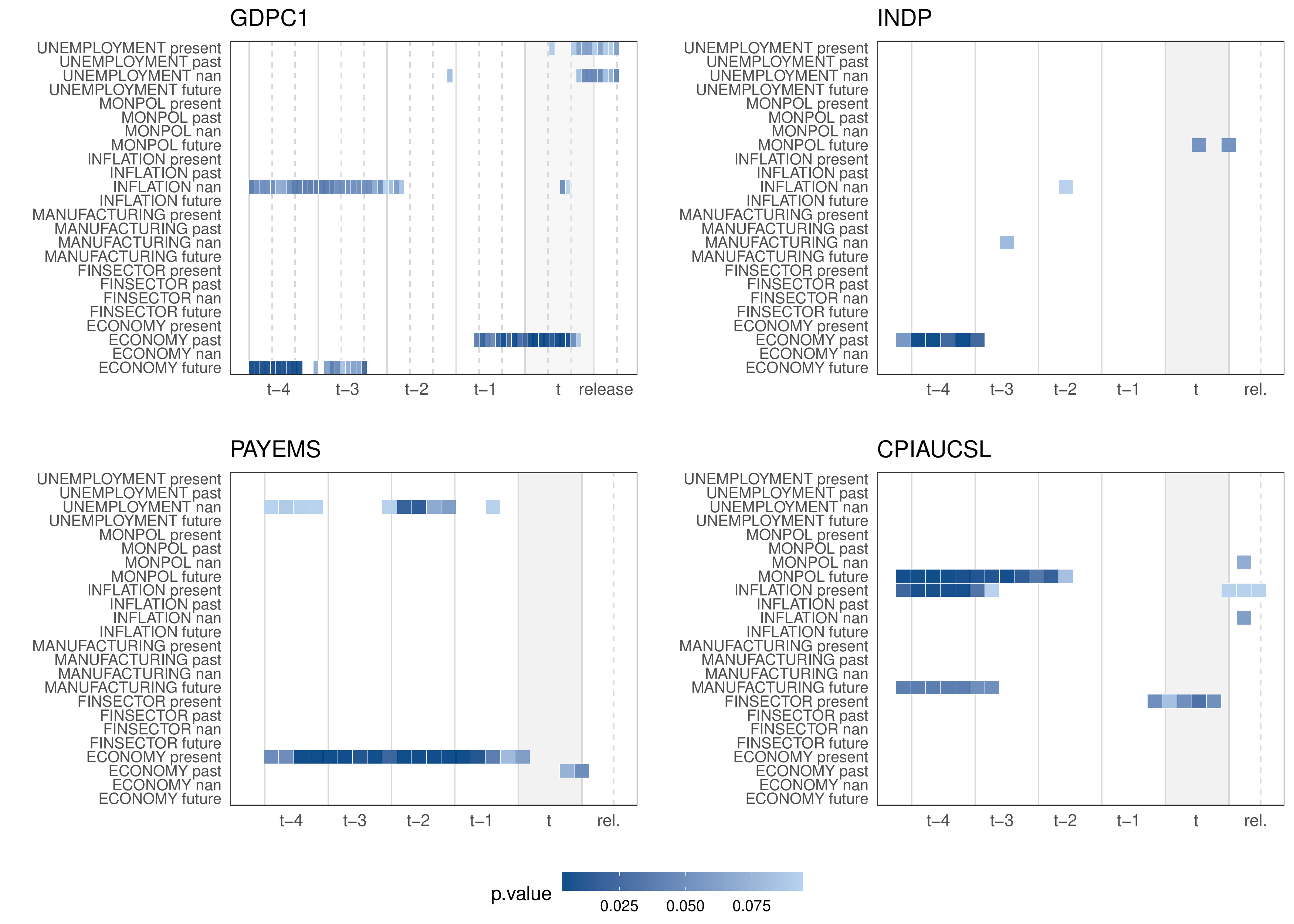}
 \caption{\small Statistical significance of the sentiment measures for all verbal tenses based on the double lasso penalized regression at each horizon $h$. 
 The colors depend on the $p$-value for those variables that are significant at least at 10\%: the darker the tile’s color, the smaller the $p$-value.
 The grey-shaded area corresponds to the nowcasting horizons.}
 \label{fig:tense_mean}
\end{figure}

Figure \ref{fig:tense_mean} shows the p-value adjusted for multiple testing of the sentiment-tense coefficient $\eta_h$ in Equation \eqref{eq:double_lasso_equation1}. 
For GDPC1 we find that at longer forecasting horizons the sentiment about the \textit{economy} is significant at the future tense, while the \textit{inflation} is relevant when it is not associated to a verbal form. When the goal is to nowcast GDP, we find that the \textit{economy} in the past tense is typically significant at 10\%, while \textit{unemployment} in the present tense shows relevance at some horizons during quarter $t$.
%% INDP
Moving to the monthly variables, the results for INDP show that at the 4-month horizon the only significant sentiment indicator is the \textit{economy} at past tense. 
%Instead, at medium and short horizons news about \textit{inflation} (in the \textit{nan} and \textit{past} tenses), \textit{manufacturing} (\textit{nan} tense), and \textit{monetary policy}, in particular for sentences with a verb in the future tense. 
When predicting PAYEMS we find that \textit{economy-present} is a statistically significant predictor across horizons from $t-4$ to $t-1$. The \textit{unemployment} indicator is also significant in predicting PAYEMS at the two and four months horizon, in particular when the verb is not detected in the sentence. Finally, the results for CPI indicate that several sentiments in the future tense are significant, in particular \textit{monetary policy} and \textit{manufacturing}, while {\it inflation} is significant at the present tense. However, at shorter horizons the present tense seems more relevant in predicting inflation for the \textit{inflation} and {\it financial sector} measures. 
Overall, when the interest is nowcasting the macroeconomic variables the results seem to suggest that sentiment measures in the past and present tenses are more often significant while the future tense and nan seem more relevant at longer horizons.

\begin{figure}[ht!]
 \centering
 \includegraphics[scale=0.45]{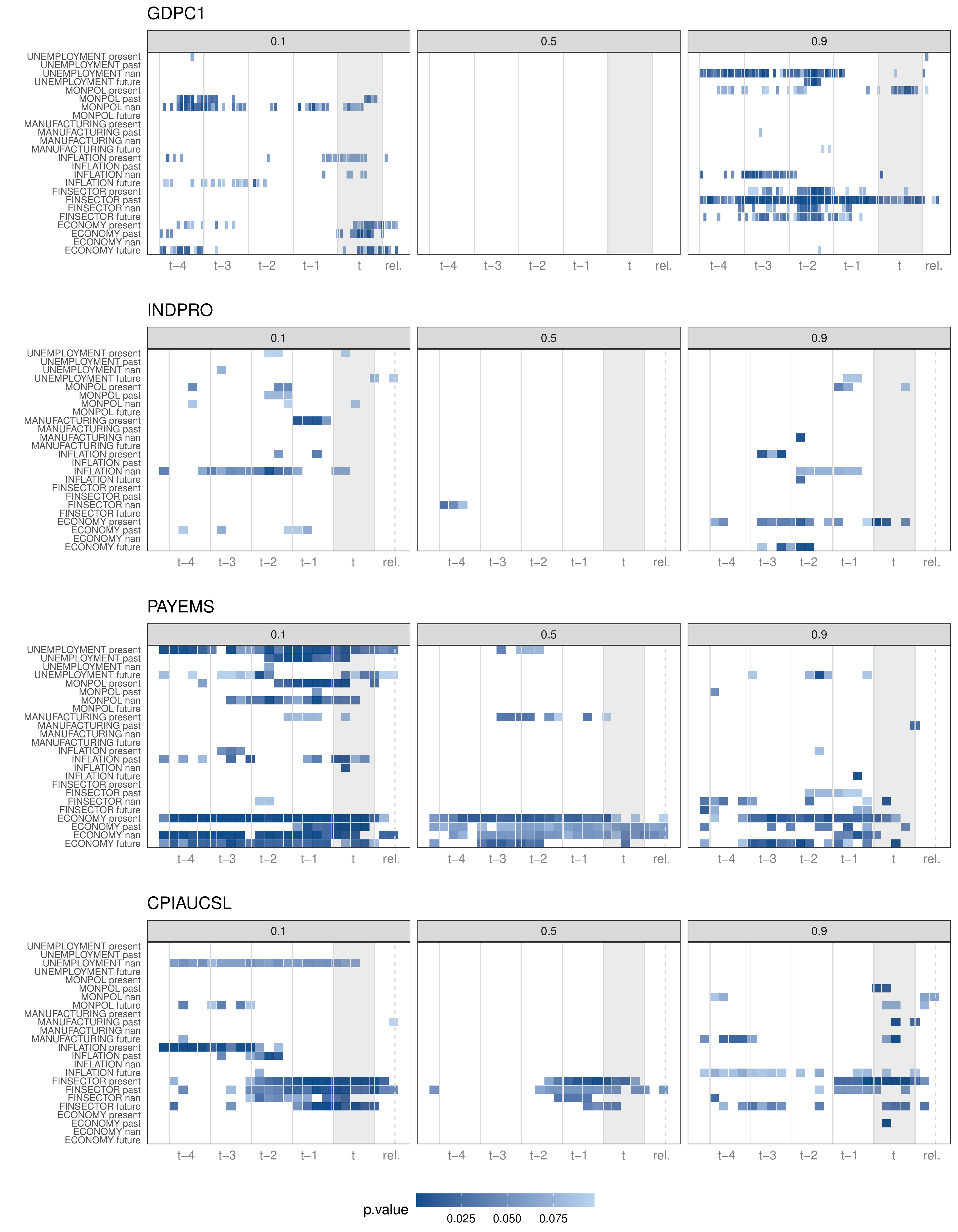} 
 \caption{\small Significance of the sentiment measures for all verbal tenses based on the double lasso penalized quantile regression at each horizon $h$ for quantiles 0.1 (left column), 0.5 (center) and 0.9 (right). The colors depend on the $p$-value for those variables that are significant at least at 10\%: the darker the tile's color, the smaller the $p$-value. The grey-shaded area corresponds to the nowcasting horizons.}
 \label{fig:tense_quantile}
\end{figure}

%% QUANTILE

We conduct the same analysis also at the quantile level and Figure \ref{fig:tense_quantile} reports the adjusted $p$-value of the sentiment coefficient in the quantile regression model in Equation \eqref{eq:double_lasso_quantile1}. Similarly to the case discussed in the paper, also splitting sentiment by verbal tense produces lower statistical significance of the measures at the median and higher significance at the tails of the distribution. In general, the results seem to suggest that the set of ToIs used to construct the sentiment indicator is a more crucial choice relative to the verbal tense. In many instances we find that several tenses for the same sentiment indicator are significant, such as in the case of predicting PAYEMS based on the \textit{economy} sentiment at the lowest quantile. Another example of this is the significance of the {\it financial sector} sentiment in predicting PAYEMS. However, there are other situations in which only one tense emerges as the most relevant, such as the \textit{economy-present} sentiment to predict the PAYEMS median and the \textit{unemployment-nan} to predict the top quantile of GDP. Interestingly, the future tense is significant when predicting GDP on the left tail when used in combination with the \textit{inflation}, {\it financial sector}, and \textit{economy} sentiment indicators.

% For variables, we observe that the proposed sentiment indicators are rarely relevant when forecasting the median, while they are often selected in the distribution tails, in particular at low quantiles, confirming that the relevance of the sentiment measures at forecasting and nowcasting growth-at-risk \citep{adrian2019vulnerable}.
%% GDP
% As it regards GDPC1, many indicators are selected at various horizons, with the \textit{financial sector} sentiment in the past and present tenses being relevant at forecasting horizons in quarter $t-2$ and $t-1$ at the 0.9 quantile.
% %% INDPRO
% Similarly to GDPC1, we observe that various indicators are selected while forecasting INDPRO, with only \textit{inflation} and \textit{economy} in the present tense appearing to be most relevant at the 0.1 and 0.9 quantiles, respectively.
% %% PAYEMS
% Moving to PAYEMS, the \textit{economy} indicators in the present and future tenses are the most selected in all three quantiles in analysis, showing quite a robust performance at both forecasting and nowcasting horizons.
% %% CPI
% Finally, the \textit{financial sector} indicators in the present and past tenses are the most selected indicators at forecasting and nowcasting horizons in $t-1$ and $t$.

\section{Comparison with the News Sentiment Index\label{sec_nsi}}

\bigskip\noindent

Measuring economic sentiment from news articles is currently an active research area \citep[among others]{thorsrud2016nowcasting, kalamaramaking, kelly2018text, bybee2019structure, shapiro2020measuring}. The main difference between the measures proposed in these papers is represented by the methodology used to extract sentiment from text. It is then an empirical question whether these measures capture genuine predictability in macroeconomic variables. 
In order to compare the performance of the FiGAS measures with these alternative sentiment indicators, we consider the News Sentiment Index (NSI) proposed by \cite{shapiro2020measuring} that is available for download at the Federal Reserve of San Francisco website\footnote{The NSI data is available at \url{https://www.frbsf.org/economic-research/indicators-data/daily-news-sentiment-index/}.}. Similarly to our sentiment, the NSI time series starts in January 1980 and it is available at the daily frequency. Table \ref{tab:cor_nsi} shows the correlations between the NSI and the six FiGAS sentiment measures at the daily, weekly, monthly, and quarterly frequencies. We consider frequencies higher than daily in order to assess the co-movement between the sentiment measures once the high-frequency fluctuations are averaged out. The results show that the correlations at the daily frequency range between 0.158 and 0.262, except in the case of \textit{inflation} where the correlation with the NSI is close to zero (this finding also extends to the other frequencies). 
Aggregating to higher frequencies, by averaging the values within the period, has the effect of increasing the level of the correlations. At the quarterly frequencies, we find that the correlations between the NSI and the FiGAS sentiments are around 0.39 for \textit{monetary policy} and \textit{manufacturing} and 0.66 for the \textit{financial sector}. Overall, it seems that the NSI has a high degree of comovement with the broader FiGAS sentiments related to \textit{economy}, \textit{unemployment} and \textit{financial sector} that is probably due to the more frequent coverage of these topics in the news.

\begin{table}[ht]
\centering
\setlength{\tabcolsep}{24pt}
\renewcommand{\arraystretch}{1.2}
\begin{tabular}[t]{lrrrr}
\hline\hline
 & Day & Week & Month & Quarter\\
\hline
ECONOMY & 0.262 & 0.439 & 0.525 & 0.581\\
FINANCIAL SECTOR & 0.251 & 0.430 & 0.559 & 0.660\\
MANUFACTURING & 0.084 & 0.197 & 0.330 & 0.392\\
INFLATION & -0.011 & -0.023 & -0.042 & -0.057\\
MONETARY POLICY & 0.158 & 0.263 & 0.319 & 0.395\\
UNEMPLOYMENT & 0.174 & 0.357 & 0.475 & 0.569\\
\hline\hline
\end{tabular}
\caption{Correlation coefficients between the NSI and the FiGAS sentiment measures. The correlations are calculated at the daily frequency and also at lower frequencies by averaging the daily values within the period.}
\label{tab:cor_nsi}
\end{table}

In order to evaluate the predictive content of the NSI and FiGAS sentiment measures, we perform an out-of-sample exercise as in the paper. We produce point forecasts from January 2002 until 2019 at horizons starting from one week before the official release of the macroeconomic variables. 
At each forecast date we select the baseline specification of the ARX model in which we use LASSO to select the most relevant regressors among lagged values of the dependent variables and current and past values of the ADS, CFNAI, and NFCI indexes. We then augment the baseline specification with the NSI and FiGAS measures and produce forecasts of the macroeconomic variables at several horizons. To evaluate the relative accuracy of the forecasts we use the aSPA multi-horizon test proposed by \cite{quaedvlieg2021multi} as discussed in the main paper. In this exercise we consider the ARXS model that uses the NSI as predictor as the null forecasts, while the FiGAS measures provide the alternative forecasts. The results of the out-of-sample comparison are provided in Table \ref{tab:oos_nsi} with $p$-values smaller than 10\% indicating that FiGAS forecasts are significantly more accurate relative to NSI across the horizons considered. 

\begin{table}[b!]
\centering
\setlength{\tabcolsep}{20pt}
\renewcommand{\arraystretch}{1.2}
\begin{tabular}[t]{lrrrr}
\hline\hline
Sentiment & GDPC1 & INDP & PAYEMS & CPIAUCSL\\
\hline
%OSS(ALL) & 0.948 & {\bf0.013} & {\bf0.009} & 0.968\\
AVERAGE & {\bf0.031} & {\bf0.004} & {\bf0.009} & {\bf0.018}\\
LASSO & 0.871 & 0.938 & {\bf0.010} & 0.929\\
ECONOMY & 0.587 & {\bf0.072} & {\bf0.009} & 0.856\\
FINANCIAL SECTOR & 0.957 & 0.182 & {\bf0.027} & {\bf0.022}\\
INFLATION & 0.405 & {\bf0.084} & 0.163 & 0.670\\
MANUFACTURING& 0.691 & {\bf0.009} & {\bf0.064} & 0.337\\
MONETARY POLICY & 0.887 & {\bf0.019} & {\bf0.062} & 0.509\\
UNEMPLOYMENT & 0.901 & {\bf0.019} & 0.781 & 0.831\\
\hline\hline
\end{tabular}
\caption{One-sided $p$-values of the aSPA multi-horizon test proposed by \cite{quaedvlieg2021multi}. The null hypothesis of the test is that the point forecasts of the ARXS model with NSI sentiment are more accurate relative to the forecasts of the ARXS that uses the FiGAS measures across the horizons considered.}
\label{tab:oos_nsi}
\end{table}

The results indicate that several FiGAS-based forecasts outperform the NSI-based for INDP and PAYEMS, as in the case of the sentiments for \textit{economy}, \textit{manufacturing} and \textit{monetary policy}. When forecasting INDP also the \textit{inflation} and \textit{unemployment} indicators contribute to provide more accurate forecasts, while {\it financial sector} is significant in the case of PAYEMS. For GDPC1 we find that the NSI-based forecasts outperform several alternative forecasts, in particular for \textit{unemployment} and {\it financial sector} while the \textit{average} of the FiGAS forecast outperforms NSI. When forecasting CPI inflation we find that the NSI and FiGAS sentiments are, in most cases, equally accurate, except in the case of the forecasts based on the {\it financial sector} sentiment that was also found in the paper to be a useful predictor of the inflation variable.

\section{Fluctuations in forecasting performance\label{sec_fluctuation}}

The out-of-sample results provided in the paper average the performance of the model forecasts over time, leaving open the question of whether any forecasting advantage derives from occasional episodes of predictability of the news sentiment, as opposed to a uniformly better performance over time. We use the fluctuation test proposed by \cite{giacomini2010forecast} to evaluate the possibility of time variation in the forecast performance.
The fluctuation test consists of performing a predictability test on rolling windows of the out-of-sample period and using appropriate critical values to account for the non-standard setting. The null hypothesis of the test is that the ARX forecasts are more accurate relative to the ARXS forecasts and negative values of the test statistic indicate rejections of the null hypothesis. We calculate the fluctuation at three representative horizons and for a rolling window of 8 quarters for GDP and 24 months for the monthly macroeconomic variables. 

\begin{figure}[ht!]
 \centering
 \includegraphics[scale=0.5]{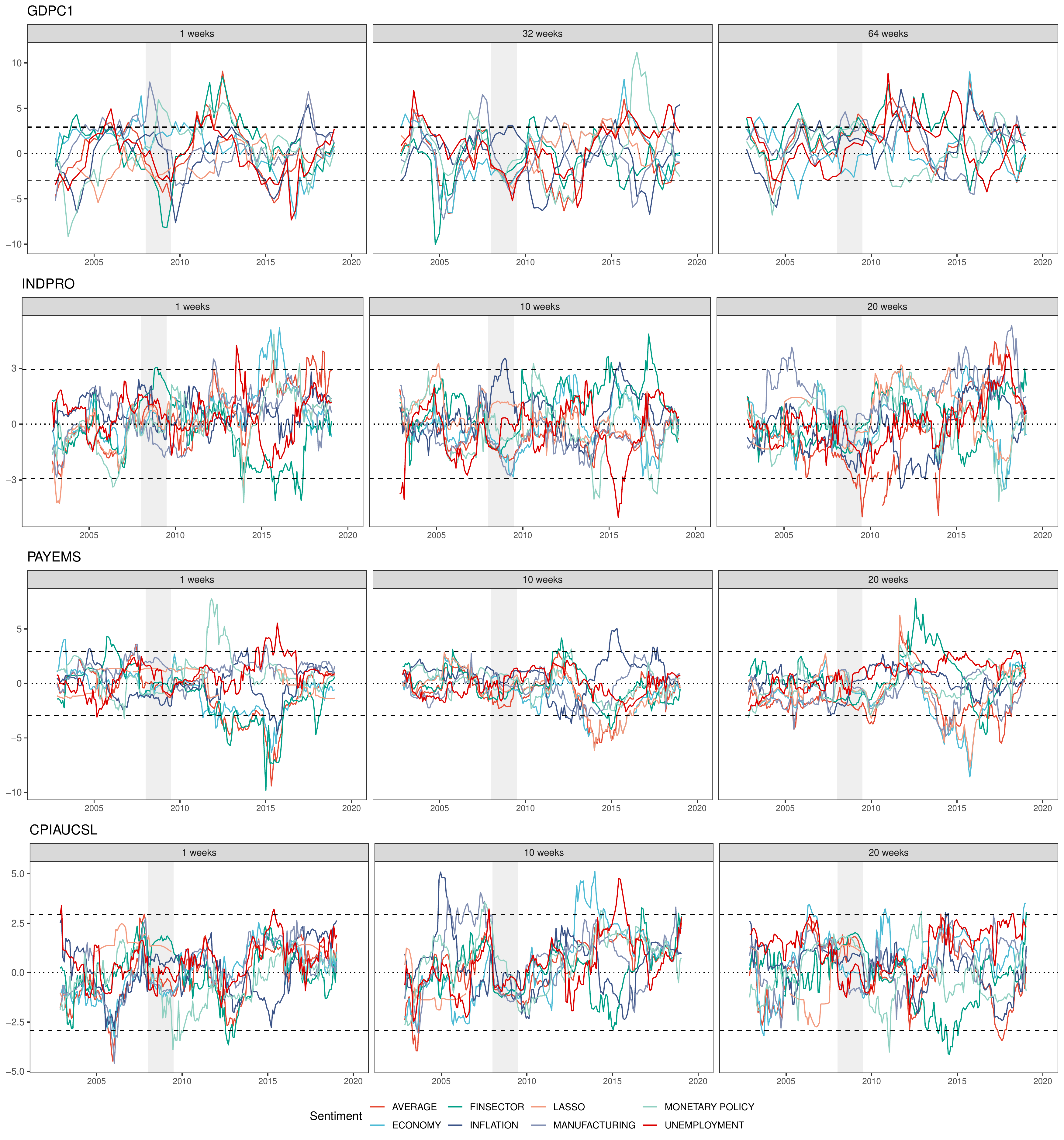}
 \caption{Fluctuation test proposed by \cite{giacomini2010forecast} with a rolling window size of 7 quarters for GDPC1 and 22 months for the monthly variables (corresponding to a 10\% window). The test statistic is aligned to the middle of the rolling window.}
 \label{fig:fluctuation_test}
\end{figure}

Figure \ref{fig:fluctuation_test} shows the results of the fluctuation test together with the one sided critical values at 10\% from \cite{giacomini2010forecast}. 
In the figure, we centered the test statistic in the middle of the rolling window so that the statistic at time $t$ represents the value of the test calculated on the window between $t-w/2$ and $t+w/2$, where $w$ represents the size of the window.
Overall, the findings suggest that sentiment derived from news delivers significantly more accurate forecasts during certain periods, but not uniformly over time. 
For instance, in the case of GDPC1 at the one and 32 week horizons some sentiment indicators provide significant predictability around 2004, during the Great Recession, and during 2015-2016. At the longest horizons the results are less remarkable, but there are still some periods with rejections of the null hypothesis of equal accuracy. Overall, it seems that the three periods discussed earlier are typically significant across the four variables considered, although not all periods for all variables. In addition, it is interesting to notice that in many occasions the predictability derives from only one or a few sentiment indicators, and only in few cases from a generalized increase of accuracy for all measures. This confirms that the six indicators capture different aspects of the economic activity which are relevant to predict some variables better than others.

\section{$R^2$ and $RMSE$ of the models\label{appendinx_R2}}

\begin{figure}[ht]
 \centering
 \includegraphics[scale = 0.6]{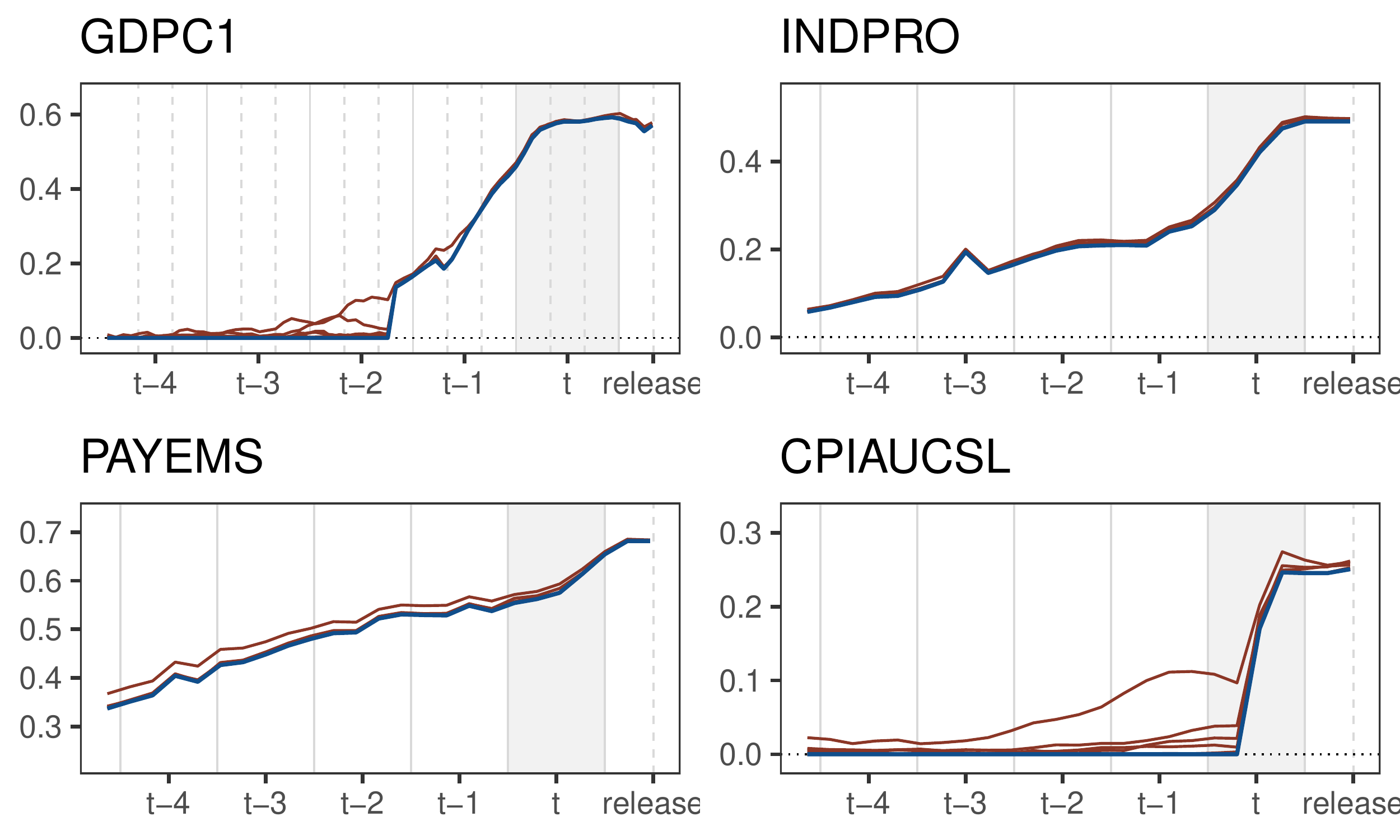}
 \caption{\small $R^2$ of ARX (blue line) and the six ARXS (red lines) models at the weekly horizons considered. The x-axis refers to quarters for GDPC1 and months for the remaining variables. The grey area represents the reference period being forecast. }
 \label{fig:rsquare_insample}
\end{figure}

\begin{figure}[ht]
 \centering
 \includegraphics[scale = 0.6]{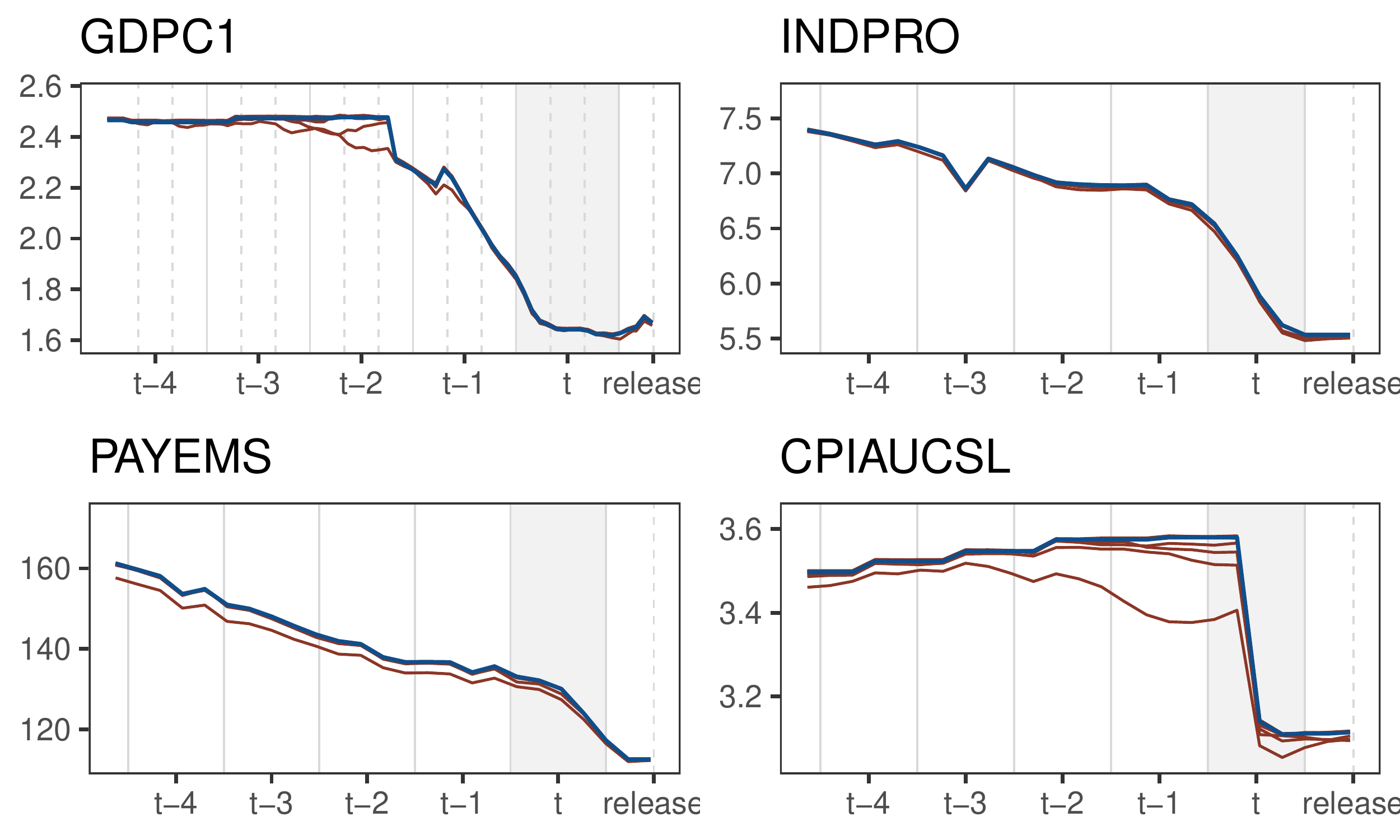}
 \caption{\small $RMSE$ of ARX (blue line) and the six ARXS (red lines) models at the weekly horizons considered. The x-axis refers to quarters for GDPC1 and months for the remaining variables. The grey area represents the reference period being forecast. The scale of the variables is percentage quarterly or monthly growth for GDPC1, INDPRO, and CPIAUCSL, while it is thousand of persons in the case of PAYEMS. }
 \label{fig:rmse_insample}
\end{figure}

Figure \ref{fig:rsquare_insample} and \ref{fig:rmse_insample} provide a descriptive analysis of the performance of the forecasting models at horizons that range from 1 day before release to four calendar periods (quarters for GDPC1 and months for the remaining variables). 
Figure \ref{fig:rsquare_insample} shows the $R^2$ goodness-of-fit statistic for the benchmark ARX forecasting model and the ARXS specification that augments ARX with each of the six sentiment indicators. The findings mirror to a large extent the findings of Figure~\ref{fig:selection_insample}, but allow to quantify in a qualitative manner the role of news sentiment measures. For GDPC1 we find that the $R^2$ of the model based on the sentiment indicators best performs during quarter $t-2$ when it is around 10.9\% higher relative to the ARX benchmark. The macroeconomic releases from quarter $t-1$ onward contribute to increase the performance of the ARX forecast while reducing the beneficial role of the news sentiment. For the monthly growth rate of the Industrial Production Index, we find that considering news increases the fit by less than 1.3\%, although we showed in Figure~\ref{fig:selection_insample} that some indicators are statistically significant. The benefit of using the {\it economy} indicator for PAYEMS is evident in the Figure, since the $R^2$ of the ARXS specification increases between 2 and 3\% at all horizons from $t-1$ to $t-4$. Finally, during months $t-2$ and $t-1$ the {\it financial sector} indicator contributes to increase significantly the goodness-of-fit, reaching a maximum increase in $R^2$ of 11\% relative to the ARX model. 

The same conclusions can be drawn by looking at the RMSE of the models. For the GDP growth rate and CPI inflation the sentiment measures contribute to reduce the RMSE up to 5.5\%, while the reductions for PAYEMS and INDPRO are smaller being up to 2.7\% and 1.2\%, respectively. 

\end{document}